\documentclass[draft]{agujournal2019}
\usepackage{url} %
\usepackage{lineno}
\usepackage{soul}
\usepackage{placeins}

\draftfalse

\journalname{JGR: Solid Earth}

\newcommand{\degree}{$^\circ$}

\begin{document}

\title{A systematic search for tectonic tremor and low-frequency earthquakes in the Atacama segment of the Chilean subduction zone (24\degree S - 31\degree S) turns up empty}

\authors{
Jannes Münchmeyer\affil{1,2},
William B. Frank\affil{2},
Sophie Giffard-Roisin\affil{1},
David Marsan\affil{1},
Anne Socquet\affil{1}
}

\affiliation{1}{Univ. Grenoble Alpes, Univ. Savoie Mont Blanc, CNRS, IRD, Univ. Gustave Eiffel, ISTerre, Grenoble, France}
\affiliation{2}{Department of Earth, Atmospheric and Planetary Sciences, Massachusetts Institute of Technology, Cambridge, MA, USA}

\correspondingauthor{Jannes Münchmeyer}{munchmej@univ-grenoble-alpes.fr}

\begin{keypoints}
\item We systematically search for low-frequency earthquakes (LFEs) and tectonic tremors in the Atacama segment in Northern Chile.
\item With classical and machine learning detection approaches, validated by manual inspection, we find no evidence for LFEs or tectonic tremors.
\item Comparing to Nankai and Cascadia, we provide constraints on the potential LFE and tremor patterns that might escape our detection.
\end{keypoints}

\newcommand{\citeneeded}{\textsuperscript{[\textcolor{blue}{citation needed}]} }

\begin{abstract}
Subduction megathrusts release stress not only seismically through earthquakes, but also through creep and transient slow deformation, called slow slip events (SSEs).
Understanding the interplay between fast and slow slip is essential for illuminating the deformation processes on the subduction interface.
The Chilean subduction margin, while one of the most seismically active regions worldwide, has few reports of SSEs.
Furthermore, there are no comprehensive reports of tectonic tremors or low-frequency earthquakes (LFEs), seismic signals typically accompanying SSEs, tracking deformation at small spatial and temporal scales.
Here, we perform a systematic search for tectonic tremors and LFEs in the Atacama segment in Northern Chile, a region hosting both shallow and deep SSEs.
Using dense seismic networks, we investigate 3.5 years between November 2020 and February 2024.
Due to the network geometry, we focus on deep tremor and LFEs.
We apply two orthogonal methods, envelope correlation for tremor search and deep learning detection for LFEs, to generate initial catalogs.
To validate the potential detections, we use clustering, matched filtering, heuristics, and extensive manual inspection.
While our initial search provides numerous candidates, after verification, we find no evidence for tectonic tremor or LFEs in the region.
In contrast, our approaches successfully recover tremors and LFEs in two reference regions outside Chile with known tremor and LFE activity.
Our observations show that tremors and LFEs in Northern Chile are either of lower moment rate than in other regions, have substantially longer recurrence rates, or are absent altogether, potentially due to the cold subduction.
\end{abstract}

\section*{Plain Language Summary}

Tectonic plates in a subduction zone accumulate stress due to tectonic forcing.
This stress is released through different deformation processes, namely earthquakes and slow slip events (SSEs).
SSEs are transient motions on the plate interface lasting from days to months releasing similar seismic moment as regular earthquakes, but without emitting seismic waves.
However, SSEs are often accompanied by seismic signals, called tremors and low-frequency earthquakes (LFEs).
So far, it is unknown if SSEs are always accompanied by tremors or LFEs or not.
In this study, we address this question for the Atacama segment in Northern Chile, one of the most seismically active regions worldwide.
We perform a systematic search for tremors and LFEs using two orthogonal methods and manual inspection.
We find no evidence for tremors or LFEs in the region.
While our results can not fully rule out the regional existence of LFEs and tremors, we provide important constraints.
We infer how big LFEs and tremors might be to still not be detectable or how rarely they need to occur.
These constraints can help in building physical models of tremorgenesis and the mechanism of SSEs, improving our understanding of the interactions between slow and fast deformation in subduction zones.

\section{Introduction}

The differential movement of tectonic plates leads to a constant accumulation of stress on the plate interface.
This stress is released either in the form of fast deformation, known as earthquakes, or through slow deformation \cite{ideScalingLawSlow2007,behrWhatThereStructures2021}.
Slow deformation can either happen constantly, in the form of creep, or transiently, as so-called slow slip events (SSE).
Slow and fast deformation have repeatedly been observed to interact with each other: 
SSEs can accelerate into fast earthquakes and earthquakes can trigger SSEs \cite{katoPropagationSlowSlip2012,radiguetTriggering2014Mw72016,vossSlowSlipEvents2018,uchidaPeriodicSlowSlip2016,obaraConnectingSlowEarthquakes2016,socquet8MonthSlow2017,itohLargestAftershockNucleation2023}.

Two seismic phenomena commonly spatiotemporally correlated with SSEs are tectonic tremors and low-frequency earthquakes (LFEs).
Tectonic tremors emit seismic signals without clear P and S arrivals that typically last several minutes \cite{obaraNonvolcanicDeepTremor2002,wechAutomatedDetectionLocation2008}.
They are called \textit{tectonic}, as they originate from an energy source on a plate interface.
For simplicity, in this study we refer to tectonic tremor simply as tremor, as we do not discuss other forms of tremor, such as volcanic tremor.
Tremors are typically observable in a frequency band between the microseismic noise ($>1$~Hz) and roughly 10~Hz.
Tremors cluster in space and time \cite{wechContinuumStressStrength2011,armbrusterAccurateTremorLocations2014}.
However, as tremors are emergent without clear phase arrivals, the definition of tremor is not sharp with no agreed definition of a minimum and maximum duration of a tremor, or the impulsiveness or smoothness of the source process.

In contrast, LFEs can be more precisely defined.
An LFE is an individual earthquake-like event on the plate interface with a source duration typically below 1~s, that differs from a regular earthquake in its spectrum, which is depleted in high-frequency content \cite{katsumataLowfrequencyContinuousTremor2003,bostockMagnitudesMomentdurationScaling2015,wangWhatMakesLowfrequency2023}.
In contrast to tremors, LFE recordings show P and S phase arrivals.
Without the high frequencies ($>$10 Hz) of earthquakes, these phases are emergent, making them substantially more difficult to pinpoint than for regular earthquakes.
In addition, LFEs are typically only observed at low signal-to-noise ratios (SNR), making it difficult to identify them visually.
LFEs are highly repetitive, with the same source producing almost identical signals hundreds or even thousands of times, called an LFE family \cite{frankUsingSystematicallyCharacterized2014,bostockMagnitudesMomentdurationScaling2015,shelly15YearCatalog2017}.
LFE families cluster in space, i.e., typically many LFE families occur within small areas.
Similarly, LFEs within and across families cluster in time with events often occurring in bursts.
This clustering is multi-scale, with very dense LFE bursts lasting tens of seconds to minutes, and then again several of such bursts occurring over a longer time frame \cite{shellyPeriodicChaoticDoubled2010,frankEvolvingInteractionLowfrequency2016}.
Interevent times between LFEs can be shorter than their wave trains, on the order of at most seconds.
It has been proposed that tectonic tremors are superposition of many closely colocated LFE sources in a short time frame \cite{shellyNonvolcanicTremorLowfrequency2007}.

Both LFEs and tremors typically coincide with SSEs \cite{itoSlowEarthquakesCoincident2007,frankUncoveringGeodeticSignature2015,michelSimilarScalingLaws2019}.
Matching LFE and tremor occurrence with geodetic records, it can be shown that they trace out areas of slow slip \cite{bartlowSpacetimeCorrelationSlip2011,wechSlipRateTremor2014,frankEvolvingInteractionLowfrequency2016,bleterySlipBurstsCoalescence2020,itohImagingEvolutionCascadia2022}.
This often leads to characteristic large scale migrations of LFE or tremor activity, with migration speeds of several kilometers per day \cite{bleteryCharacteristicsSecondarySlip2017}.
It is unclear whether tremors and LFEs are triggered by aseismic deformation \cite{frankDailyMeasurementSlow2019} or if tremors and LFE facilitate the migration of slow slip \cite{itohSliptremorInteractionVery2024}.
In addition, it has been shown that the scaling between tremors/LFEs and SSEs does not follow a purely linear relationship \cite{frankDailyMeasurementSlow2019,mouchonSubdailySlowFault2023}.
Nonetheless, tremors and LFEs are a useful tool to illuminate SSE dynamics, such as slip extent or intermittence, in particular, at scales that are too short or small to be observed geodetically \cite{frankSlowSlipHidden2016,mouchonSubdailySlowFault2023}.

SSEs, tremors, or LFEs have been observed in diverse subduction zones, for example, in Nankai, Alaska, Cascadia, Mexico, Costa Rica, Chile, Hikurangi and Taiwan \cite{behrWhatThereStructures2021}.
The types and number of detections vary substantially by region, likely through a combination of different activity rates and different levels of instrumental density.
In Nankai, hundreds of SSEs have been reported, accompanied by countless tremor and LFE detections \cite{nishikawaReviewSlowEarthquakes2023}.
In addition, the region hosts very low-frequency earthquakes, a related class of events with much lower frequency content ($<0.1$~Hz).
In contrast, some regions, like the Atacama segment in Northern Chile, only have a handful of SSE detections and no comprehensive LFE or tremor observations \cite{kleinDeepTransientSlow2018,kleinReturnAtacamaDeep2022}.
Most reports of deep tremors and LFEs come from hot subduction zones, yet some cold subduction zones like Hikurangi host deep tremor as well \cite{behrWhatThereStructures2021,aden-antoniowLowFrequencyEarthquakesDowndip2024}.
It has been proposed that hot subduction favors tremorgenesis as larger amounts of fluids are release in metamorphic reactions at shallower depth than in cold subduction zones \cite{conditSlabDehydrationWarm2020,behrWhatThereStructures2021}.

In subduction zones hosting SSEs, tremors, and LFEs, the spatial distribution of these events is heterogeneous with variations along both strike and dip.
Along dip, two bands, separated by the seismogenic region of the locked plate interface, typically these events: a deep band (down to 60~km) and a shallow band (up to 10~km) \cite{safferFrictionalHydrologicMetamorphic2015}.
There are more reports of deep SSEs, tremors, and LFEs than shallow ones, yet this might be observational bias as shallow events typically occur further offshore than their deep counterparts and the density of seafloor geophysical instruments is low \cite{safferFrictionalHydrologicMetamorphic2015,nishikawaSlowEarthquakeSpectrum2019}.
In this study, we use onshore instrumentation and therefore focus on deep tremor and LFEs.
However, our methods are designed sufficiently general to detect shallow events as well, yet with lower sensitivity.

Similar to the along dip variation, SSE activity varies along strike, such as differences in recurrence times in Cascadia \cite{michelSimilarScalingLaws2019}.
The observed spatial distribution of tremors and LFEs if even more heterogeneous, with events clustering in nests of just few kilometers in diameter \cite{armbrusterAccurateTremorLocations2014}.
At the same time, it can not be ruled out that geodetic slip shows similar heterogeneities at small scales, as these would be below the geodetically resolvable limits.

In this study, we focus on the Northern Chile subduction zone between 24\degree S and 31\degree S (Figure~\ref{fig:station_map}).
This region is a known seismic gap with no major earthquake ($M>8$) since at least 1922 \cite{vignySearchLostTruth2024}.
At the center of this region, in the area of the subducting Copiapó ridge, shallow and deep SSEs have been detected.
The deep slow slip has a recurrence time of 5 to 6 years with good instrumental records for the 2014 and 2020 events and limited coverage for the 2009 event \cite{kleinDeepTransientSlow2018,kleinReturnAtacamaDeep2022}.
The only geodetically well-resolved shallow SSE occurred in 2023 and was accompanied by a seismic swarm \cite{ojedaSeismicAseismicSlip2023,munchmeyer2024chile_sse}.
As the region hosts repeating seismic swarm activity \cite{marsanEarthquakeSwarmsChilean2023}, and seismic swarms have previously been linked to SSEs \cite{nishikawaRecurringSlowSlip2018,gardonioRevisitingSlowSlip2018}, it is likely that the area hosts recurrent SSEs \cite{ojedaSeismicAseismicSlip2023,munchmeyer2024chile_sse}.
No tremors or LFEs accompanying the SSEs have been reported.
Therefore, the region of the Copiapó ridge and its surroundings is a prime candidate for searching for tremors and LFEs, as they would be expected in the vicinity of SSEs.
We focus our study on deep tremor and LFEs, as the lack of offshore instrumentation will reduce sensitivity for shallow tremor.
Nonetheless, we do not exclude shallow tremor from our search.

While LFEs and tremors are likely to share a source process \cite{shellyNonvolcanicTremorLowfrequency2007}, their different waveform features lead to different challenges in detecting and characterising them.
Tremors, with their duration of several minutes, are usually easier to identify than individual LFEs, whose low SNR waveforms only last for a few seconds.
At the same time, the lack of clear arrivals means that tremor location methods rely on low-frequency waveform coherence, with methods such as backprojection and envelope correlations.
This leads to high location uncertainties, in particular, in terms of hypocentral depth.
This problem is further aggravated because tremors sources cannot be adequately modeled by point sources.
In contrast, LFEs have identifiable P and S arrivals that can be used for locating their sources.
However, due to their low SNR waveforms, it is often difficult to identify the arrivals of an individual LFE phase with sufficient precision.
To overcome this challenge, the waveforms of similar LFE sources can be stacked together to increase the SNR of their waveforms allowing for picks that can be more accurately determined.
Such similar LFE sources are typically identified through template matching.
While their precise definition and the possibility to identify more accurate locations makes LFEs more powerful to understand the structure and dynamics of plate motion on the subduction interface \cite{mouchonSubdailySlowFault2023}, the short duration of LFEs makes them substantially harder to detect than tremors.
Therefore, in this search for tremors and LFEs in Chile, we use orthogonal methods targeting their specific characteristics.
We will first present our results for tremors as finding tremor candidates is potentially easier, followed by the search for LFEs.

\section{Data and Methods}

\subsection{Study region and data availability}

\begin{figure*}[ht!]
\centering
\includegraphics[width=\textwidth]{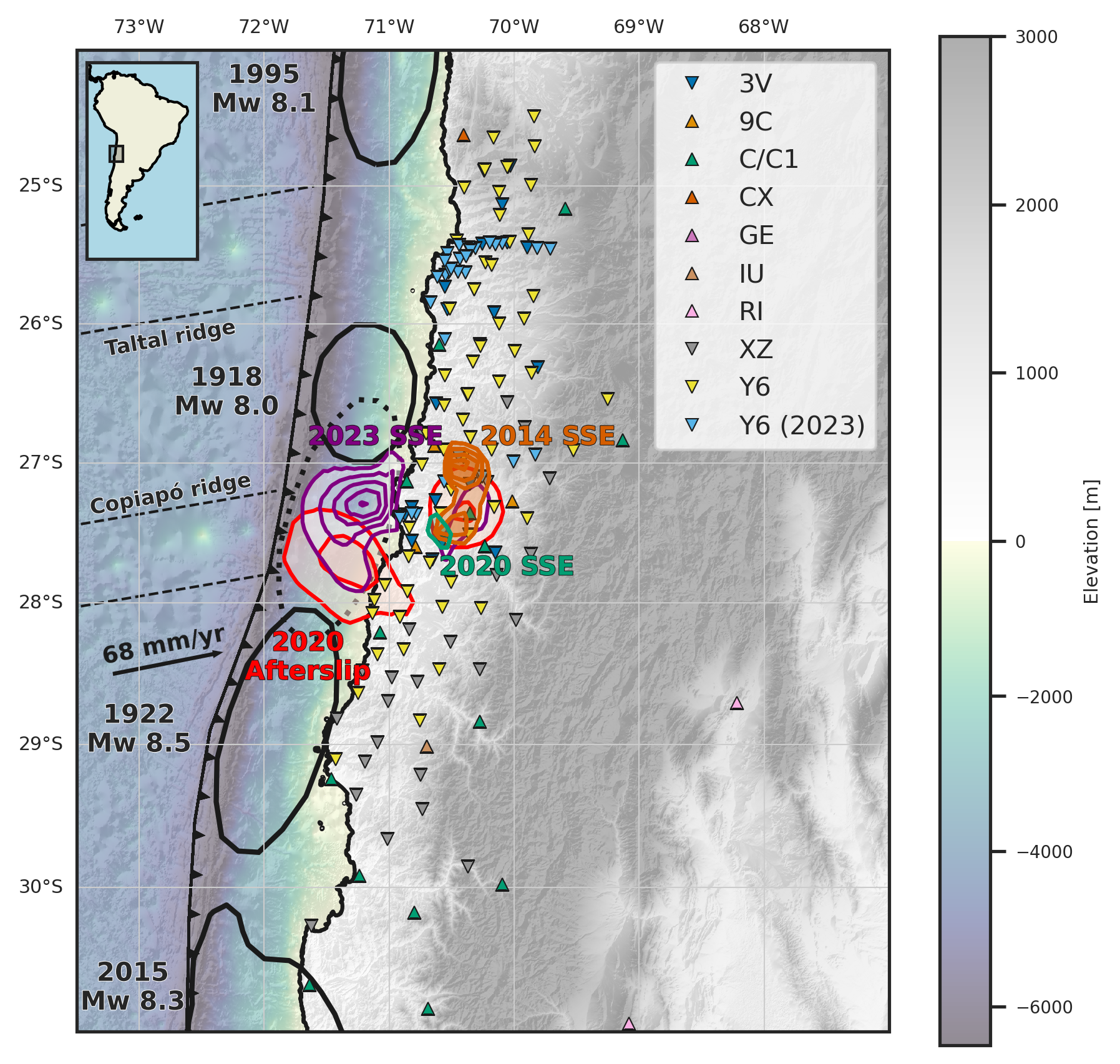}
\caption{Overview map of the region visualizing the tectonic setting, the past slow slip activity, and the seismic station coverage. Stations are colored according to their seismic network, using upward triangles for permanent stations and downward triangles for temporary stations. Within the Y6 network, stations deployed for four months in 2023 are indicated separately. A small number of additional stations of the C1, CX, GE and RI networks outside the visualised area have been used. Colored contours show past SSEs. For the 2014 SSE we provide the 5~cm slip contours \cite{kleinDeepTransientSlow2018}, for the 2020 SSEs the approximate slip outline \cite{kleinReturnAtacamaDeep2022}, for the 2020 afterslip the 2~cm slip contours \cite{molina2024sse}, and for the 2023 SSE the 1~cm slip contours \cite{munchmeyer2024chile_sse}.}
\label{fig:station_map}
\end{figure*}

Our study region is the Atacama segment of the Northern Chile subduction zone between 24\degree S and 31\degree S (Figure~\ref{fig:station_map}).
The region is known to host a wide range of slow and fast seismic phenomena.
The Atacama segment hosts one of the longest standing seismic gaps along the Chilean megathrust, called the 1922 gap, after the 1922 $M_w \approx 8.5$ earthquake \cite{kanamoriNewConstraints19222019,vignySearchLostTruth2024}.
The segment is bordered in the North by the 1995 Antofagasta earthquake ($M_w = 8.0$) and to the South by the 2015 Illapel earthquake ($M_w = 8.3$).
Between these two past rupture zones, the subducting Copiapó ridge is thought to have repeatedly served as a boundary or nucleation region for major earthquakes \cite{ruizHistoricalRecentLarge2018}.
Recent research found that the ridge also served as a boundary for the 1922 earthquake \cite{vignySearchLostTruth2024}, suggesting that the ridge acts as a persistent seismic barrier.
This is consistent with the low degree of interseismic locking ($<0.6$) inferred from continuous GNSS records \cite{metoisInterseismicCouplingMegathrust2016,yanez-cuadraInterplateCouplingSeismic2022,gonzalez-vidalRelationOceanicPlate2023}.
\citeA{munchmeyer2024chile_eqs} provide a detailed description of the seismicity in the area.

In addition to its role in the nucleation and arrest of megathrust ruptures, the subducting Copiapó ridge is the only area in our study region known to host SSEs.
Both shallow and deep SSE have been reported in the area.
For shallow SSEs, only one certain detection exists, occurring in September 2023 \cite{munchmeyer2024chile_sse}.
Another shallow SSE in 2006 has been proposed, yet geodetic station coverage is insufficient to fully verify this detection \cite{ojedaSeismicAseismicSlip2023}.
The shallow SSE in 2023 had a magnitude $M_w=6.6$ and lasted for approximately one month \cite{munchmeyer2024chile_sse}.
This SSE was accompanied by migrating seismic swarm activity illuminating the rupture process.
In 2020, a seismic sequence with cumulative moment $M_w=7.1$ triggered afterslip propagating northwards outside the rupture area for more than 50~km \cite{molina2024sse}.
Similar to the 2023 SSE, this afterslip was accompanied by seismic warms.
This suggest that seismic swarms mark transient slow slip, similar to observations in other regions \cite{nishikawaRecurringSlowSlip2018}.
As swarm activity has repeatedly been observed during the past 50 years, it has been proposed that the region hosts recurring shallow SSEs of different sizes \cite{comteSeismicityStressDistribution2002,marsanEarthquakeSwarmsChilean2023,ojedaSeismicAseismicSlip2023,munchmeyer2024chile_sse}.
Further downdip, deep SSEs with a recurrence period of $\sim$5~years and a typical duration of more than one year have been observed \cite{kleinDeepTransientSlow2018,kleinReturnAtacamaDeep2022}.
For the best-studied event in 2014, a magnitude of $M_w=6.9$ and a duration of more than a year have been reported \cite{kleinDeepTransientSlow2018}.
With moment rates around $3 * 10 ^ {12}$~Nm/s (shallow) and $6 * 10 ^ {11}$~Nm/s (deep), these SSEs are at the lower end of typical tectonic SSE moment rates \cite{ideSlowEarthquakeScaling2023}.
There is one report of tectonic tremor in the region of the Copiapó ridge \cite{pasten-arayaAlongDipSegmentationSlip2022}.
Yet these tremors are only occur on a single day and do not coincide with an SSE.
We revisit this report in the discussion in section~\ref{sec:pasten_araya}.

Here, we use a dense seismic network comprised of 193 seismic stations (triangles in Figure~\ref{fig:station_map}).
The stations are a combination of permanent networks and temporary deployments and include broadband sensors, short-period instruments, and geophones.
We focus on the time range between the deployment of the first temporary stations in November 2020 and the removal of the last stations in February 2024.
As not all networks have been running at the same time, the typical number of available stations per day is between 60 and 120.
The stations consist of a mix of broadband instruments, short-period seismometers, and geophones.
Each of these instruments provides adequate coverage for the typical tremor band (1~Hz to 10~Hz).
A more detailed description of the seismic networks is provided in \cite{munchmeyer2024chile_eqs}, where the dataset is used to build a dense catalog of regular earthquakes.

\subsection{Reference regions with known tremor/LFE activity}

To test the sensitivity of our methods in identifying LFEs and tremors from continuous waveforms, we apply our workflows to two reference regions with known tremor and LFE activity, Cascadia and Nankai.
We focus on an intense period of intense tremor and LFE activity in each region.
Compared to the study region in Chile, we consider smaller spatial extents and time periods in the two reference regions.
This is sufficient for the sensitivity test and substantially reduces computational demand.
We analyze the differences in station density and seismic noise conditions of all three regions in the discussion.

The first reference region is Southern Vancouver Island, Cascadia, Canada.
The region has long been known to host SSEs \cite{dragertSilentSlipEvent2001}, tectonic tremors \cite{wechAutomatedDetectionLocation2008}, and LFEs \cite{bostockMagnitudesMomentdurationScaling2015}.
For this region, we study the time period between 15/01/2021 and 15/02/2021, as during this period a tremor migration occurred underneath the study region.
We use all stations from the Canadian seismic network between 48.0\degree N and 49.3\degree N and between 125.9\degree W and 122.7\degree W.
In total, we use 13 seismic stations.

The second reference region is Nankai, Japan.
Similar to Cascadia, there are extensive reports of SSEs, tectonic tremors, and LFEs \cite{shellyLowfrequencyEarthquakesShikoku2006,katoDetectionDeepLowfrequency2020,takemuraReviewShallowSlow2023}.
We use data from HiNet, a dense network of borehole seismometers.
Similar to Cascadia, we selected a time period with known tremor activity, in this case from 25/05/2012 to 15/06/2012.
We focus on the region between 32.8\degree N and 34.5\degree N and between 132\degree E and 134\degree E, using 36 HiNet stations.

\subsection{Tremor and LFE detection workflows}

\begin{figure*}[ht!]
\centering
\includegraphics[width=\textwidth]{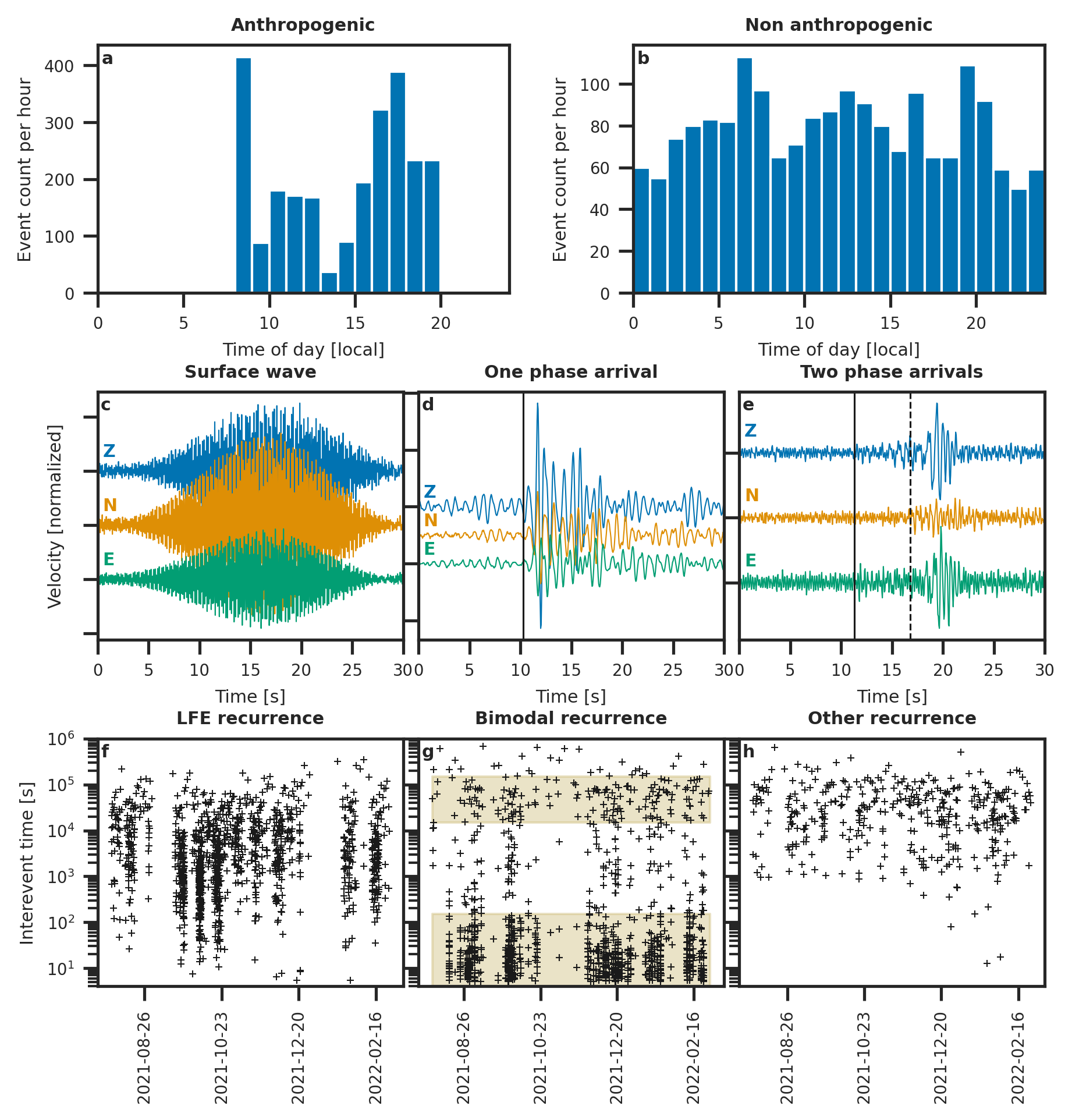}
\caption{The three features used for LFE classification with typical examples. \textbf{a-b} Distribution of events over time of day, identifying anthropogenic and non-anthropogenic sources. \textbf{c-e} Wave type and number of identifiable phase arrivals. Note that for the actual classification, we are basing our assessment on all available stations and not a single station. The three colors show the three components with a vertical offset for better visibility. \textbf{f-h} Recurrence time plots with log-scaled interevent times on the y axis. All examples in the figure are from candidate event families in Northern Chile. In g, we highlight the two typical recurrence time bands.}
\label{fig:lfe_classification}
\end{figure*}

We developed workflows for tremor and LFE detection combining different classical and machine learning methods with manual inspection.
Here, we provide a general overview of these workflows, while we provide all additional technical detail in \ref{sec:method_detail}.

The basis for our tremor detection is the envelope correlation method of \cite{wechCatalogingTectonicTremor2021}.
For this method, we calculate the signal envelopes in the tremor band (1 to 8~Hz) at each station, a proxy for the seismic moment rate.
We correlate the envelopes between stations in 5~minute windows to identify spatially coherent signals.
If the envelopes across multiple stations are well correlated, we estimate differential travel times from the cross-correlations and locate events using a grid search.
Due to the size of the study region, we subdivide the region into multiple grid cells.
From these initial detections, we remove all detection coinciding with cataloged earthquakes \cite{munchmeyer2024chile_eqs}, as these likely correspond to earthquake rather than tremor sources.
As tremors occur in bursts, we use clustering with DBScan and only keep events in clusters, thereby removing spurious detections \cite{wechCatalogingTectonicTremor2021}.

For LFE detection, we adapt a classical earthquake detection workflow consisting of phase picking, phase association, and location.
We use the deep learning LFE phase picker from \citeA{munchmeyerDeepLearningDetects2024}.
We associate the picks using PyOcto \cite{munchmeyerPyOctoHighthroughputSeismic2024}.
We locate the events with NonLinLoc \cite{lomaxProbabilisticEarthquakeLocation2000} using the 3D velocity model from \citeA{munchmeyer2024chile_eqs}.
The low SNR waveforms of LFEs makes the analysis of individual LFEs challenging, so we use all detected LFE candidates as a basis for template matching.
This way, we generate a family of similar detections for each candidate event.
This allows us to analyse the group behaviour of the family and the stacked waveform, which have a better SNR than individual detections.
Similar to the tremor detection described above, we perform the template matching in multiple space-time cells.
As it is likely that closely located candidate templates will detect similar families of events, we perform a deduplication among the template results using the temporal behaviour of their event families.

We manually inspect the remaining clusters to analyze three diagnostic features (Figure~\ref{fig:lfe_classification}).
First, we identify whether the occurrence times throughout the day suggest an anthropogenic source, e.g., mining, or a non-anthropogenic source (Figure~\ref{fig:lfe_classification}a,~b).
Second, we identify if the signal shows no clear phase arrivals, typical for a surface wave, one phase arrival, or two phase arrivals (Figure~\ref{fig:lfe_classification}c-e).
Third, we classify if the recurrence pattern of the event family exhibits the typical temporal clustering of LFEs \cite{shellyPeriodicChaoticDoubled2010,frankEvolvingInteractionLowfrequency2016}, has a bimodal recurrence, or another type of random behaviour like the Poissonian interevent distribution of regular earthquakes (Figure~\ref{fig:lfe_classification}f-h).
We expect an LFE to be non-anthropogenic, show two phase arrivals, and have a typical LFE recurrence pattern.
To ensure we do not miss LFE families, we err on the side of caution and consider a candidate event family to be LFE-like if in doubt.
For example, we assign two phase arrivals if at least some station stacks show two clear arrivals.

\section{Results}

\subsection{Tremor candidates}

\begin{figure*}[ht!]
\centering
\includegraphics[width=\textwidth]{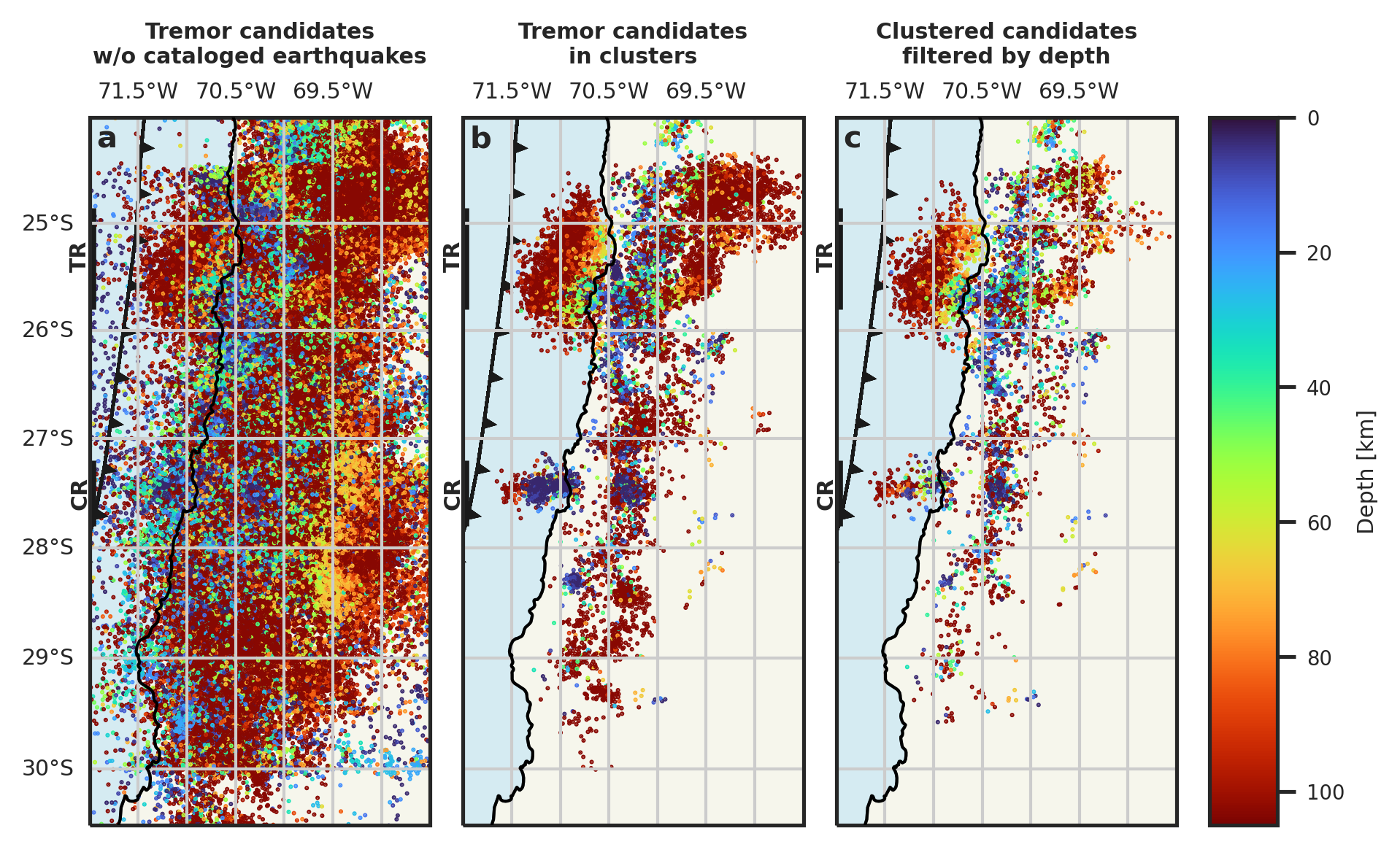}
\caption{Tremor candidates in Northern Chile detected using envelope correlation. \textbf{a} Detected tremor candidates excluding detections matching cataloged earthquakes. \textbf{b} All tremor candidates contained in DBScan clusters with at least 5 members. \textbf{c} Same as b but without tremor candidates in shallow ($<10$~km) and deep ($>95$~km) clusters. We note that none of the candidates have been identified as an actual tectonic tremor. In all plots, we add a small scatter (Gaussian with standard deviation 0.03\degree) to the tremor locations to reduce overlap caused by the quantization of the grid search (4~km). We indicate the approximate locations of the incoming Copiapó (CR) and Taltal (TR) ridges.}
\label{fig:tremor_overviews}
\end{figure*}

\begin{figure*}[ht!]
\centering
\includegraphics[width=\textwidth]{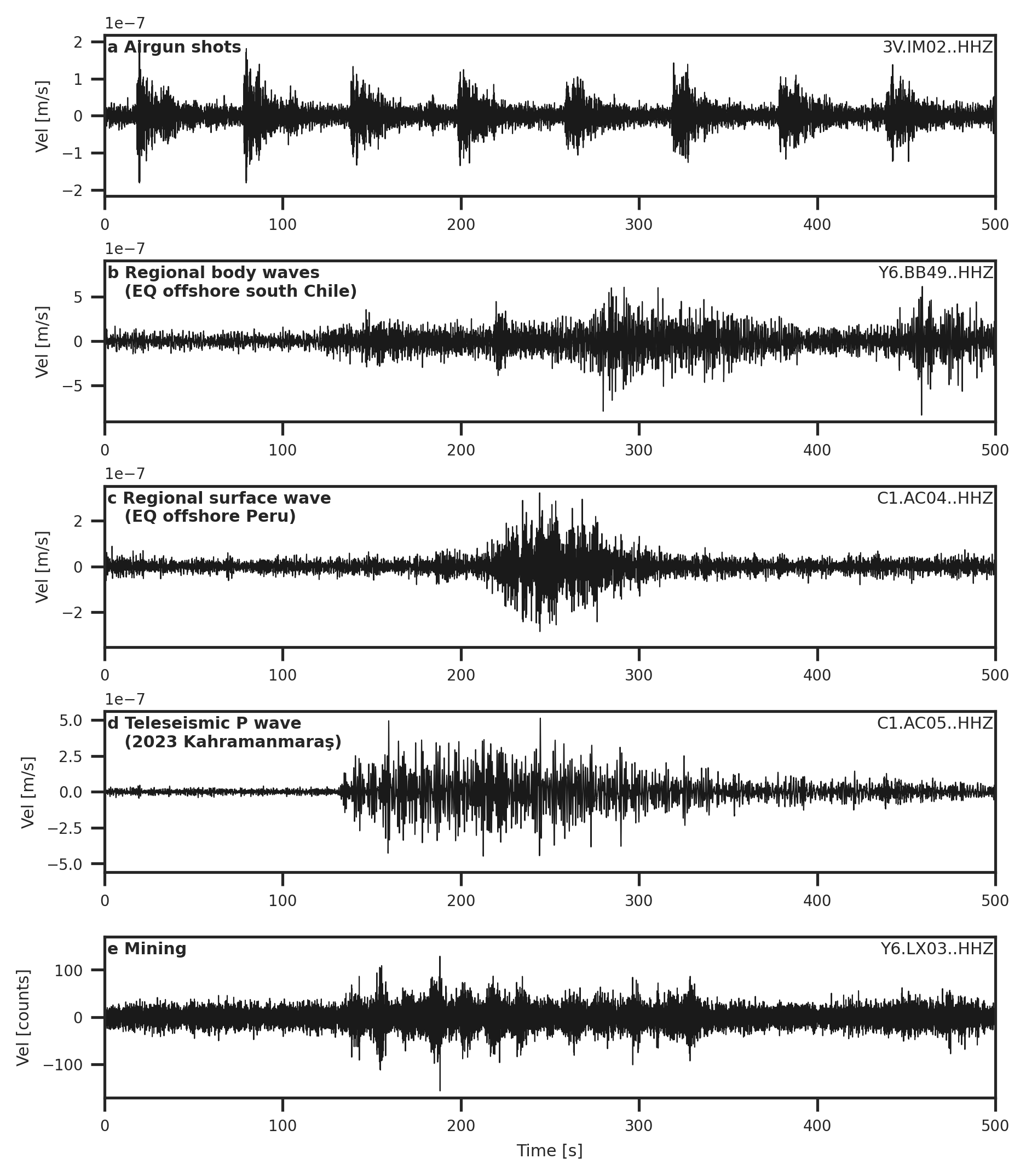}
\caption{Vertical-component waveforms from signals identified as potential tremors. All waveforms have been bandpass-filtered between 1 and 10~Hz, the typical tremor frequency band. Labels on the top right indicate the SEED identifier of the trace. Note that we classified these waveforms not solely based on the single station waveform, but incorporating data from further stations, wider frequency bands, and additional context information.}
\label{fig:tremor_examples}
\end{figure*}

To identify tremor candidates, we apply the envelope correlation to the data from our 3.5 year study period, identifying 430,000 tremor candidates in total.
Note that many candidates will have been counted multiple times due to the overlapping search regions.
Of these detections, 230,000 coincide with cataloged earthquakes, leaving 200,000 candidates (Figure~\ref{fig:tremor_overviews}a).
These tremor candidates are distributed throughout the whole study region with an along-dip extent between 200~km and 300~km.
The distribution loosely mirrors the station coverage, i.e., the along dip extent is wider in regions with wider station coverage.

The majority of detections occur isolated in space, time, or both.
After applying DBScan and only retaining clusters with at least 5 members, only 12,816 tremor candidates remain in 1156 clusters.
Of these clusters 147 have an average depth shallower than 10~km, suggestive of surface sources such as mining activity.
447 clusters have an average depth deeper than 95~km, which is close to the lower boundary of the search region set at 100~km.
These events are substantially deeper than expected for tremors and are likely caused by deep or teleseismic sources.
We exclude both these very shallow and very deep clusters, leaving 562 clusters.
We show selected waveform examples from these clusters in Figure~\ref{fig:tremor_examples}.

Of the 562 clusters, 44 occur offshore around 25.5\degree S within three weeks between 07/03/2023 and 27/03/2023.
These detections correspond to the airgun shots of an active seismic experiment within the cruise SO297 of the RV Sonne \cite{lange2023high}.
The waveforms of these detection show a very regular 60~s periodicity (Figure~\ref{fig:tremor_examples}a).
Figure~\ref{fig:cruise_tremors} shows that these detections trace out the track of the ship during the cruise.
This highlights that our tremor detection approach has better resolution in latitude direction than in longitude direction, because the events are occurring offshore outside the network.
We do not continuously detect the airgun shots, missing detections in particular when the ship is far offshore.
For all events associated with the cruise, we overestimate the depth and locate the events too close to shore.
This is likely caused by the large azimuthal gap and the assumption of an S wave moveout, even though the P phases of the airgun shots dominate their waveforms.

After removing the cruise-related clusters, we manually inspect all remaining clusters.
For each cluster, we visually inspect the waveforms from the stations with high-correlation in the envelope correlation in the typical tremor frequency band of 1-10 Hz.
For all but 42 clusters, we identify clear impulsive onsets in the waveforms, showing that the signals are not tremors.
The onsets seem to result from two sources: mine blasting and regional earthquakes.
For mine blasting, we recognized many events with strong surface wave content, which have not been captured completely in the earthquake catalog.
The average depth in these mining related clusters is typically just below the 10~km threshold set earlier, meaning they were not automatically removed by this criterion.
These events can reach sufficient cluster size, as mine blasts are often performed in sequences.
In addition, as for all event types, the same blast will often be visible in different overlapping detection regions, increasing the apparent number of events in a cluster through duplicate detections.
The regional earthquakes often feature S minus P times in excess of 30~s.
This means that the candidates are either not within the study area of the earthquake catalog used for comparison, or their location is too far from the inferred tremor locations to be matched in our automated exclusion procedure.
Many of these regional events originate from the highly active and deep Jujuy cluster around 24\degree S \cite{valenzuela-malebranSourceMechanismsRupture2022}.
For regional earthquakes, we often observe multiple consecutive tremor detections because of the overlapping windows and high envelope correlation in the coda of large events.
Together with the overlap of the detection cells, this can fulfill the DBScan clustering criteria.

We analyse the remaining clusters by inspecting the 15 minutes around the tremor candidates.
In addition to the 1 to 10~Hz band, we study the 10 to 20~Hz band to identify impulsive onsets at higher frequencies.
Through this process, we identify 33 further clusters as earthquake-related.
Some of the other clusters are caused by high envelope correlation from either the coda wave of local earthquakes or the surface wave train of regional earthquakes (Figure~\ref{fig:tremor_examples}b,~c).
We also see detections caused by early aftershocks, where the impulsive onsets were hard to identify in the 1 to 10~Hz band because of the high background noise but clearly visible in the 10 to 20~Hz band.
One cluster can be directly related to the arrival of the P wave from the 2023 Kahramanmaras earthquake (Figure~\ref{fig:tremor_examples}d), possibly related to the source duration and directivity of that event \cite{jia2023complex}.
Particularly challenging to identify were clusters with mixed causes, e.g., where the first detection is spurious while subsequent members of the cluster contain earthquake signals.

Of the remaining 9 clusters, we identified 4 clusters that correspond to mining in the mines Las Cenizas, Las Luces, and Franke (Figure~\ref{fig:tremor_examples}e).
These clusters are characterised by locations close to the mines, regularly repeating impulses, and clusters during the daytime.
The majority of the detected locations in these clusters are very shallow, however, they did not meet the rejection criterion for shallow clusters due to a small number of deep outlier detections.

Three clusters occur in close proximity to each other on 2021-12-11 between 2 pm and 3 pm (UTC).
Upon close inspection, these clusters are caused by a mixture of three sources.
First, we suspect mine blasting caused some of the detections.
Second, some detections are caused by the teleseismic P wave arrivals from a magnitude 5.1 earthquake in the South Sandwich Islands with origin time at 14:35:44 UTC.
Third, at 14:54:19 UTC a magnitude 5.7 earthquake occurred about 40~km north of La Serena, causing numerous detections in its coda.
The combination of these three sources produced clusters that could not be excluded with the criteria set out above.

\begin{figure*}[ht!]
\centering
\includegraphics[width=\textwidth]{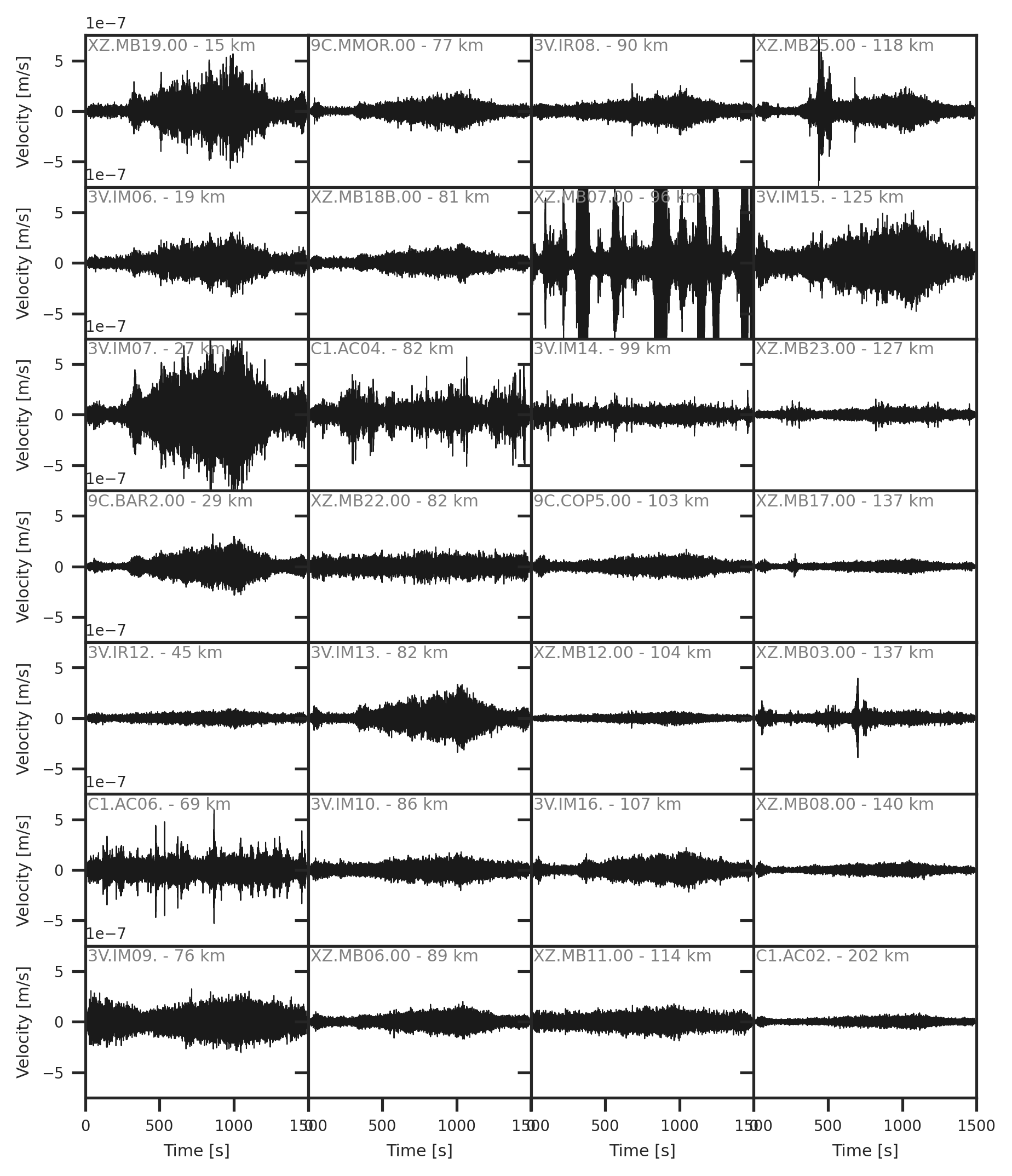}
\caption{Emergent 20~min signal on 2023-06-23, likely related to a marine or seafloor process. The inferred location is not compatible with plate boundary faulting like a tremor. Each panel shows the vertical component waveforms for a station between 26.5\degree S and 28.5\degree S for the time from 2023-06-23 21:22:00 UTC to 2023-06-23 21:47:00 UTC. Spectrograms for the same stations and time range are available in Figure~\ref{fig:472_spectrogram}. We removed the instrument response and bandpass filtered the waveforms between 1 and 8~Hz. All stations are scaled equally. Stations are ordered by their epicentral distance to the most likely source of the 20~minute signal estimated with envelope correlation (Figure~\ref{fig:472_enveloc}).}
\label{fig:472_waveform}
\end{figure*}

The remaining two clusters occur on 2023-06-23 between 21:20 and 21:50 (UTC) and on the subsequent day between 04:00 and 06:00 (UTC), both at a latitude around 27.5 degrees and near the coast.
The first cluster consists of numerous detections within an extended signal of twenty minute duration.
The signal is roughly cigar shaped with a slow increase in amplitude over 15 minutes and a subsequent decay over 5 minutes (Figure~\ref{fig:472_waveform}) with a peak amplitude of about $5 * 10^{-7}$~m/s.
The signal is strongest between about 3 and 10~Hz, with stations near the source showing signal up to 20~Hz (Figure~\ref{fig:472_spectrogram}).
At the closest station, MB19, additional even higher frequency burst are visible.
At low-noise stations, the signal can clearly be identified even at a distance of 200~km.

For further inspection, we locate the signal using a modified envelope correlation.
In contrast to the original location using a 5 minute window, we use a 20 minute correlation window, thereby making use of the full signal shape.
The most likely event location is close to shore at a depth of 3~km, the most shallow depth considered in the envelope correlation (Figure~\ref{fig:472_enveloc}).
Considering the uncertainty distribution, a shallow depth ($<$20~km) is likely.
For the epicenter, the uncertainty ellipse trends South-Westward, away from the most dense part of the network.

The events 7 hours later are similar in terms of shape and frequency content, but each individual signal is shorter, only about 200~s long (Figures~\ref{fig:477_waveform},~\ref{fig:477_spectrogram},~\ref{fig:477_enveloc}).
The peak amplitudes are about 4 times lower.
While these shorter and lower amplitude signals are harder to locate, the most likely locations are close to the location of the preceding 20~minute signal.
The combination of similar signal characteristics and close temporal proximity suggests that both clusters are caused by the same source process.

Based on the observed characteristics, we suggest the signals are not caused by tectonic tremors.
First, the clusters occur in isolation within the 3.5 year time period, which does not fit the typical recurrence of any known tremor in other regions.
Second, the content of high frequency signals is rather strong for a tremor, in particular the near-source signals above 20~Hz.
Third, while the source depth has substantial uncertainty, it is without question shallower than the plate interface that locates at almost 40~km depth in this region \cite{munchmeyer2024chile_eqs}.
In addition, at interface depth, the epicentral region of the potential tremor shows high levels of interface seismicity \cite{munchmeyer2024chile_eqs}.
This is incompatible with both shallow and deep tremor sources, which typically occur updip/downdip of the most active band of interface seismicity.
While we are not able to uniquely identify a source process for the signal, we suggest that it is occurring on or near the ocean floor.
A potential candidate would be a turbidity current in the area below the outlet of the Copiapó river where a sediment fan has been mapped \cite{warwelSeismicStructureTectonics2025}.
Similar seismic signatures have been observed for other turbidity current events, including the clustered occurrence, \cite{bakerSeabedSeismographsReveal2024} and for mass wasting events on land \cite{cookDetectionPotentialEarly2021,clareSeismicAcousticMonitoring2024}.

After this manual inspection of the initially detected tremor candidates, no candidates remain.

\subsection{LFE candidates}

\begin{figure*}[ht!]
\centering
\includegraphics[width=\textwidth]{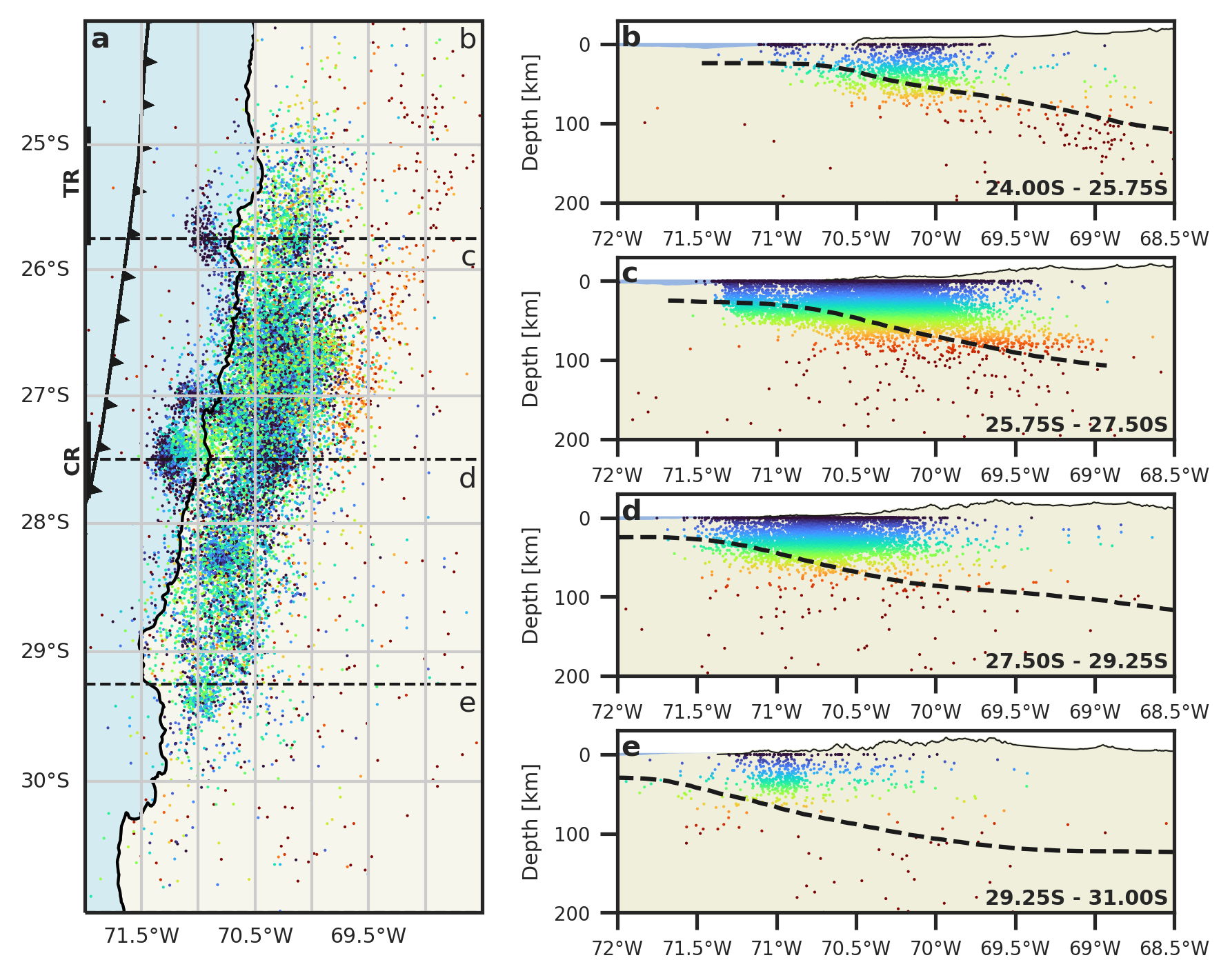}
\caption{LFE candidates in Chile detected using the deep learning pipeline \cite{munchmeyerDeepLearningDetects2024}. Each dot represents an individual LFE candidate colored according to the depth. There is a small number of detections at larger depths and outside the selected map region. The dashed lines in \textit{a} delineate the different cross-sections in b to e. The dashed black lines in b to e indicates the slab model from \citeA{munchmeyer2024chile_eqs} at the center of the profile. We show the topography at the center of the profile. Topography is exaggerated by a factor of 4, bathymetry is not exaggerated. We indicate the approximate locations of the incoming Copiapó (CR) and Taltal (TR) ridges. We note that we identify none of the LFE candidates as a true detection.}
\label{fig:chile_lfes_raw}
\end{figure*}

\begin{figure*}[ht!]
\centering
\includegraphics[width=\textwidth]{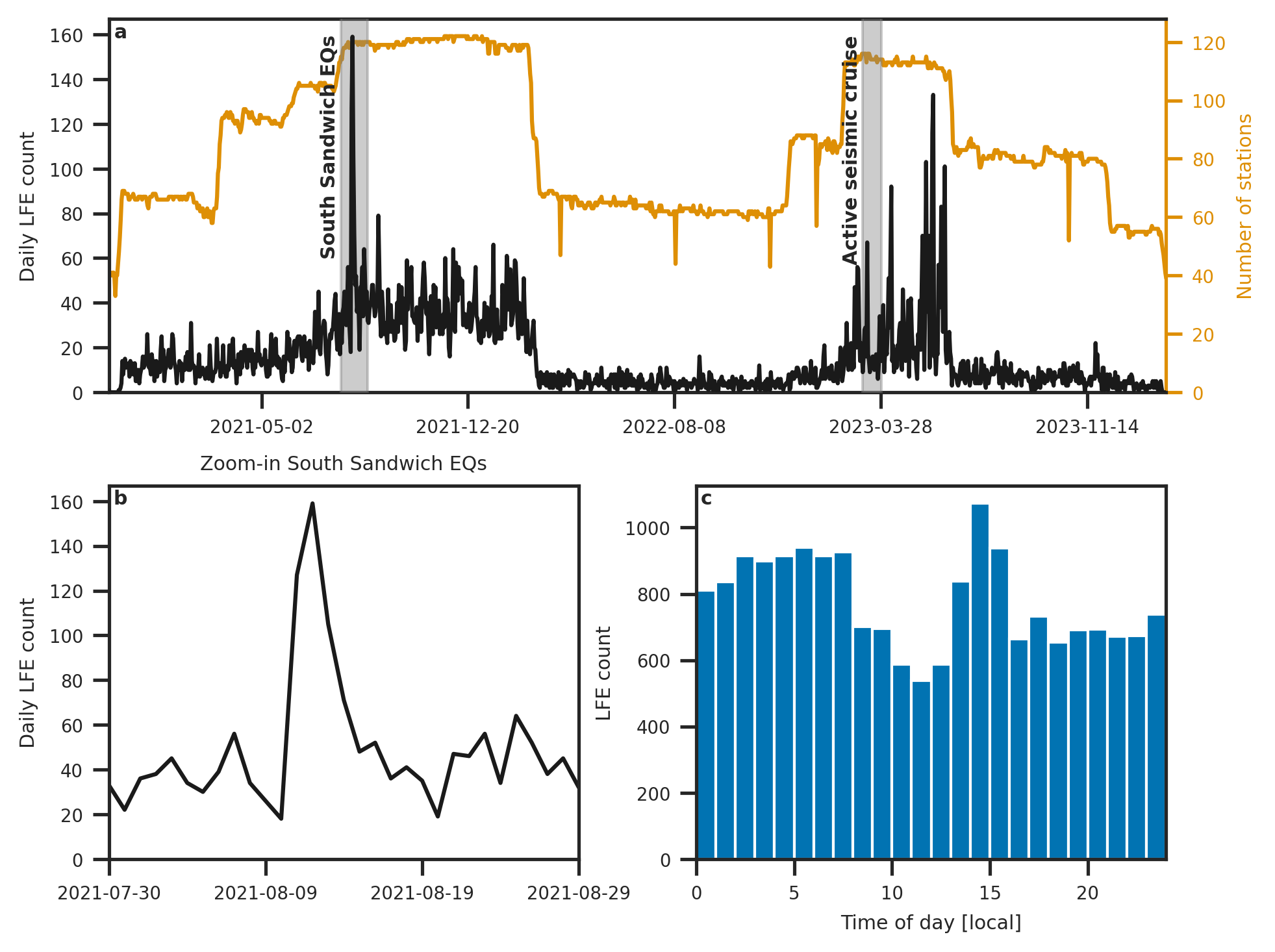}
\caption{Temporal behaviour of the LFE candidates detected in Chile. \textbf{a} In black, daily LFE count throughout the study period. In orange, daily number of active seismic stations. \textbf{b} Zoom-in of panel a around the South Sandwich EQ sequence. \textbf{c} Hourly LFE candidate count by time of day.}
\label{fig:chile_lfes_timing}
\end{figure*}

\begin{figure*}[ht!]
\centering
\includegraphics[width=\textwidth]{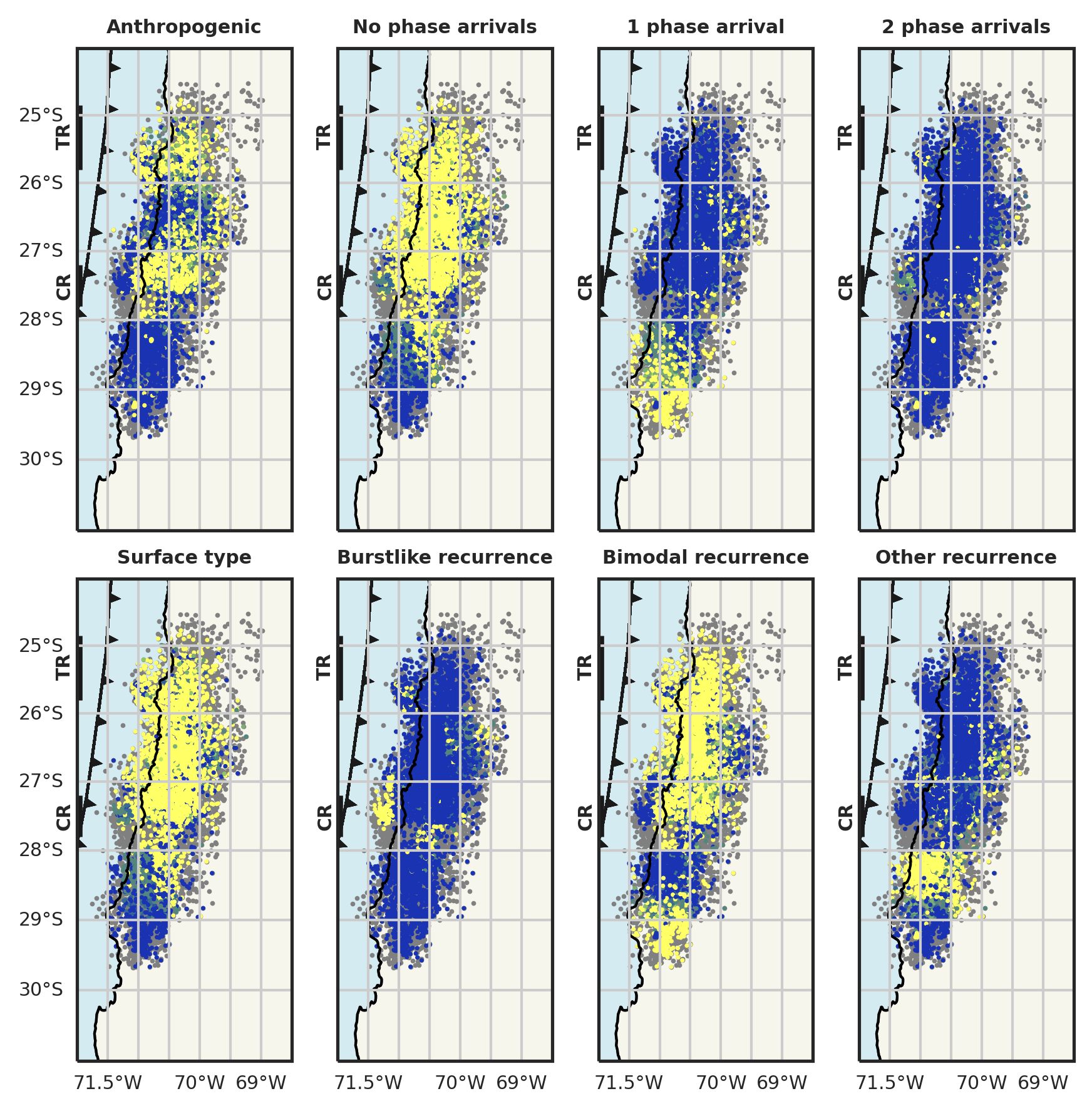}
\caption{Summary of the classification results of LFE candidate families after template matching. Each panel visualises one feature. Each dot marks one template, i.e., one LFE candidate from the initial deep learning detections. The colors indicate whether the template fits the criterion not at all (blue) or fully (yellow). In-between values are possible, because a template might have been classified differently between the different grid cells and time blocks it belongs too. Grey dots have been added for templates that have not been manually inspected because they did not match with any cluster or no member of the cluster had a sufficient number of detections. We indicate the approximate locations of the incoming Copiapó (CR) and Taltal (TR) ridges.}
\label{fig:chile_lfes_classification_summary}
\end{figure*}

\begin{figure*}[ht!]
\centering
\includegraphics[width=\textwidth]{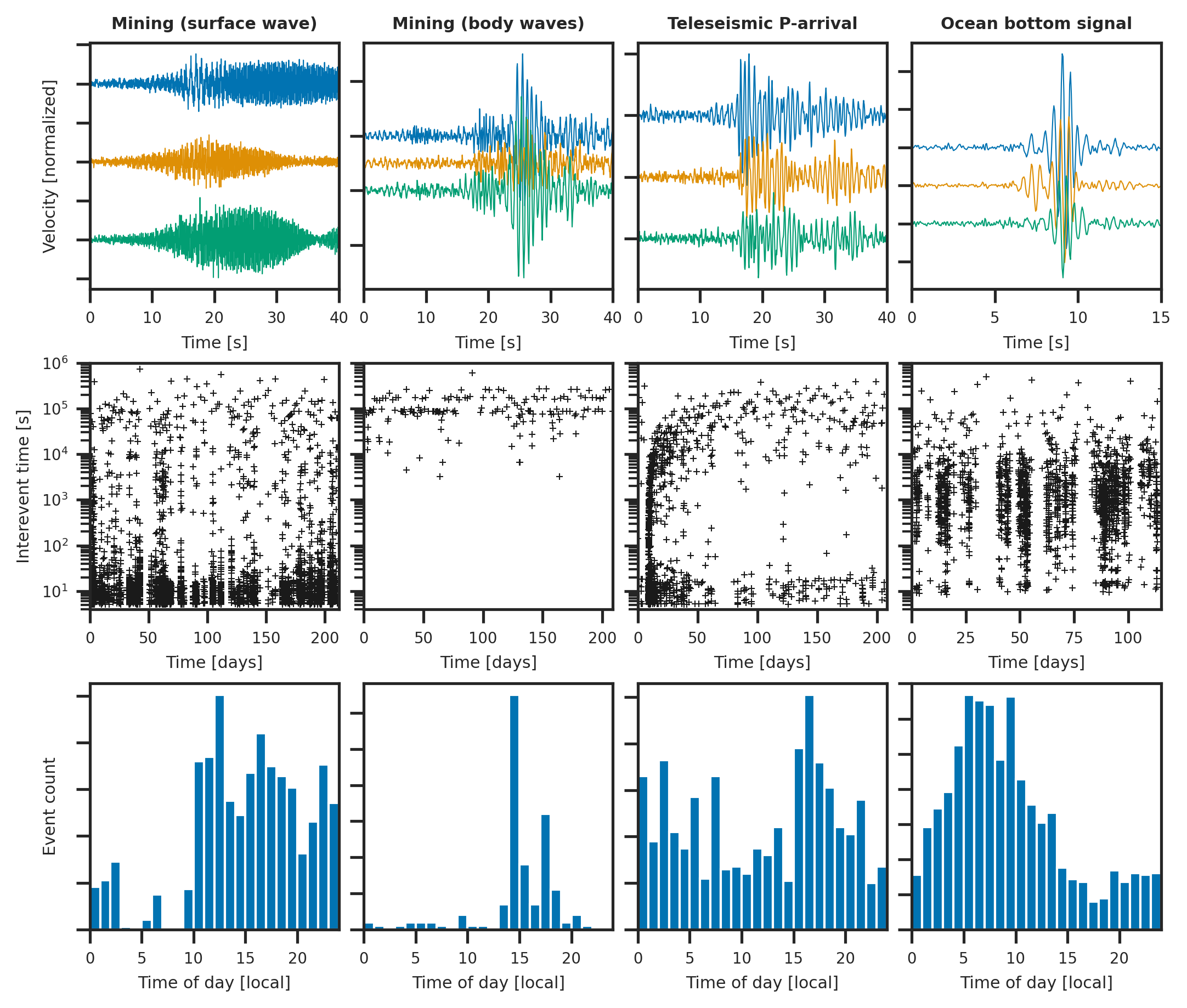}
\caption{Waveforms (top), recurrence plots (center) and time of day histograms (bottom) of potential LFE families. Each column represents one family, with a label on top. The waveform plots show ZNE components (top to bottom). Note that we only show a selected station per family, while we used multiple stations for classification. The recurrence plots show the interevent time between two consecutive detections. The time-of-day histrograms have been normed for visibility, while event counts differ between the examples. The teleseismic P-arrival family corresponds to earthquakes in the South-Sandwich islands, with the aftershock sequence of the $M_w=8.1$ earthquake on 12th August 2021 clearly visible in the recurrence plot (middle).}
\label{fig:lfe_examples}
\end{figure*}

Using a deep learning pipeline developed to detect LFEs within continuous seismic waveforms \cite{munchmeyerDeepLearningDetects2024}, we find a total of 18,640 LFE candidates in the study area (Figure~\ref{fig:chile_lfes_raw}).
On average, each event has 13.6 associated picks, with an almost equal split into P and S picks.
The epicenters are distributed primarily along a band of about 100~km East-West extent starting slightly east of the coastline and ranging from 25\degree S to 29.5\degree S.
Two shallow offshore clusters are detected around 25.5\degree S and 27.5\degree S.
The majority of detections locates between the surface and 60~km depth.
Detections at larger depths only occur systematically in the northeastern portion of the study region.
No clear structures are visible in the epicentral or hypocentral distributions, however, the detection method is known to produce high picking uncertainties and thereby large location uncertainties, particularly in depth \cite{munchmeyerDeepLearningDetects2024}.

Looking at the temporal evolution of the detection rate, we see that its first order behavior is determined by the number of available stations (Figure~\ref{fig:chile_lfes_timing}).
The highest average number of detections occurs between August 2021 and March 2022, the time during which both the XZ and Y6 networks were deployed (Figure~\ref{fig:station_map}).
Another time span of high detection rates occurs in early 2023, the time when two dense networks of geophones were deployed to record offshore active seismic experiments.
Detection rates during this time are potentially further increased by false detections related to the airgun shots.
In comparison to periods of intense tremor activity in regions where LFEs and tremor regularly occur, the detection rates we find in northern Chile are low \cite{munchmeyerDeepLearningDetects2024}.
In addition, the dynamic range, i.e., the ratio between days with low activity and days with high activity with similar network coverage, is lower than in \cite{munchmeyerDeepLearningDetects2024}, suggesting a high base rate of false detections.

We observe a clear spike in event counts on 12th August 2021 that slowly decreases over the next days to weeks (Figure~\ref{fig:chile_lfes_timing}).
The increased activity occurs throughout the study region and is not localised.
Through waveform inspections and comparison to global seismic catalogs, we identified this spike as related to teleseismic activity around the South Sandwich Islands.
Beginning with the 12th August 2021 $M_w=7.5/8.1$ earthquake doublet \cite{jia2021SouthSandwich2022}, a large number of detections occurs within P waves of earthquakes beneath the South Sandwich Islands (Figure~\ref{fig:chile_lfes_timing}b).
The decay in apparent LFE activity following the mainshocks correlates well with the decay in aftershock activity.
We identified aftershocks of the South Sandwich sequence down to $M_w \approx 5.0$.
Upon further manual inspection, we identified further detections related to other teleseismic P arrivals, e.g., from events offshore Japan.
Nonetheless, these detections alone do not account for all detections.

Figure~\ref{fig:chile_lfes_timing}c breaks down the activity by time of day to identify potential anthropogenic sources of the detections.
There is a general trend for higher detections at night, pointing at higher detection capabilities due to lower nighttime noise.
However, there is a clear spike in activity in the daytime between 1 and 4~pm local time.
This spike in activity in the early afternoon is consistent with the typical blasting schedule of open pit mines in the area \cite{munchmeyer2024chile_eqs}, suggesting that some detections can be attributed to mine blasting.

To further investigate the detected candidates, we perform template matching and manual classification, inspecting 581 potential LFE families in total (Figure~\ref{fig:chile_lfes_classification_summary}).
Figure~\ref{fig:lfe_examples} shows typical event classes detected.
A large number of families can be related to anthropogenic sources (Figure~\ref{fig:lfe_examples} first and second columns).
Many of these are characterised by surface type waveforms.
Furthermore, most families show a bimodal recurrence pattern, i.e., they have two characteristic return periods.
This pattern is common for heavy machinery that is operated temporarily, in particular, in the open pit mines.
This differs from the burst-like recurrence of typical LFE families as these typically show intermediate return times as well.
Only few families show burst-like recurrence, including a number of families around the Copiapó ridge.
Looking at the number of phase arrivals, the most common case is the complete absence of clearly identifiable arrivals.
This is consistent with the high number of surface wave type examples.
In particular in the South, we detect a substantial number of events with a single arrival.
For many of these detections, the amplitudes are highest on the vertical component, suggesting that these families might be related to teleseismic P wave arrivals (Figure~\ref{fig:lfe_examples} third column).
For some of these families, we identified the source region of the teleseismic waves, with a particularly high occurrence of waves from the South Sandwich islands region.
Only a handful of all families feature two clear phase arrivals.
No family jointly fulfills all criteria we set for an LFE detection, i.e., non-anthropogenic, two phase arrivals, and burst-like recurrence.

While no family can be clearly identified as an LFE, we acknowledge that the distinction between burstlike activity and bimodal activity is subjective.
Therefore, to further rule out LFE detections, we manually pick P and S phase arrivals for all events with two arrivals and locate these events.
We skip some examples originally identified as having two phases if we could not pick reliable P wave arrivals for at least two stations.
All of these excluded examples were strongly dominated by the later arriving Rayleigh wave, showing that they correspond to very shallow sources.
In addition, we skip two examples with P to S times above 30~s at the closest stations as these would locate either very deep or far outside the network, not reasonable locations for LFEs.

Of the located families, the majority falls into the Los Colorados mine (28.3\degree S, 70.8\degree W) or the Candelaria mine (27.5\degree S, 70.3\degree W).
Two families locate around 27.48\degree S 70.97\degree W at 14~km depth about 7~km offshore (Figure~\ref{fig:lfe_examples} last column).
One family locates around 26.94\degree S 70.86\degree W at only 0.1~km depth about 6~km offshore.
Comparing the locations of these families to the slab model and seismicity from \citeA{munchmeyer2024chile_eqs}, these events locate significantly shallower than the interface, even taking the depth uncertainties (8 to 13~km) into account.
Notably, the locations are also shallower than the most shallow regular seismicity detected in these regions.
In contrast, given the location uncertainties, the events might be located on the shoreline and at the surface/seafloor.
This is also consistent with the waveforms, which show strong surface waves.
All three families show higher event rates during the daytime (early morning to early afternoon), even though we did not label them as anthropogenic in our original classification, because they show activity at all times.
This suggests anthropogenic origins of the signals, potentially related to fishery or other boating.
Regardless of their exact origin, these signals and their locations are not consistent with LFE sources.
The location analysis therefore confirms that none of the candidate detections have typical LFE characteristics.

\subsection{Tremors and LFE detections in the reference regions}

\begin{figure*}[ht!]
\centering
\includegraphics[width=\textwidth]{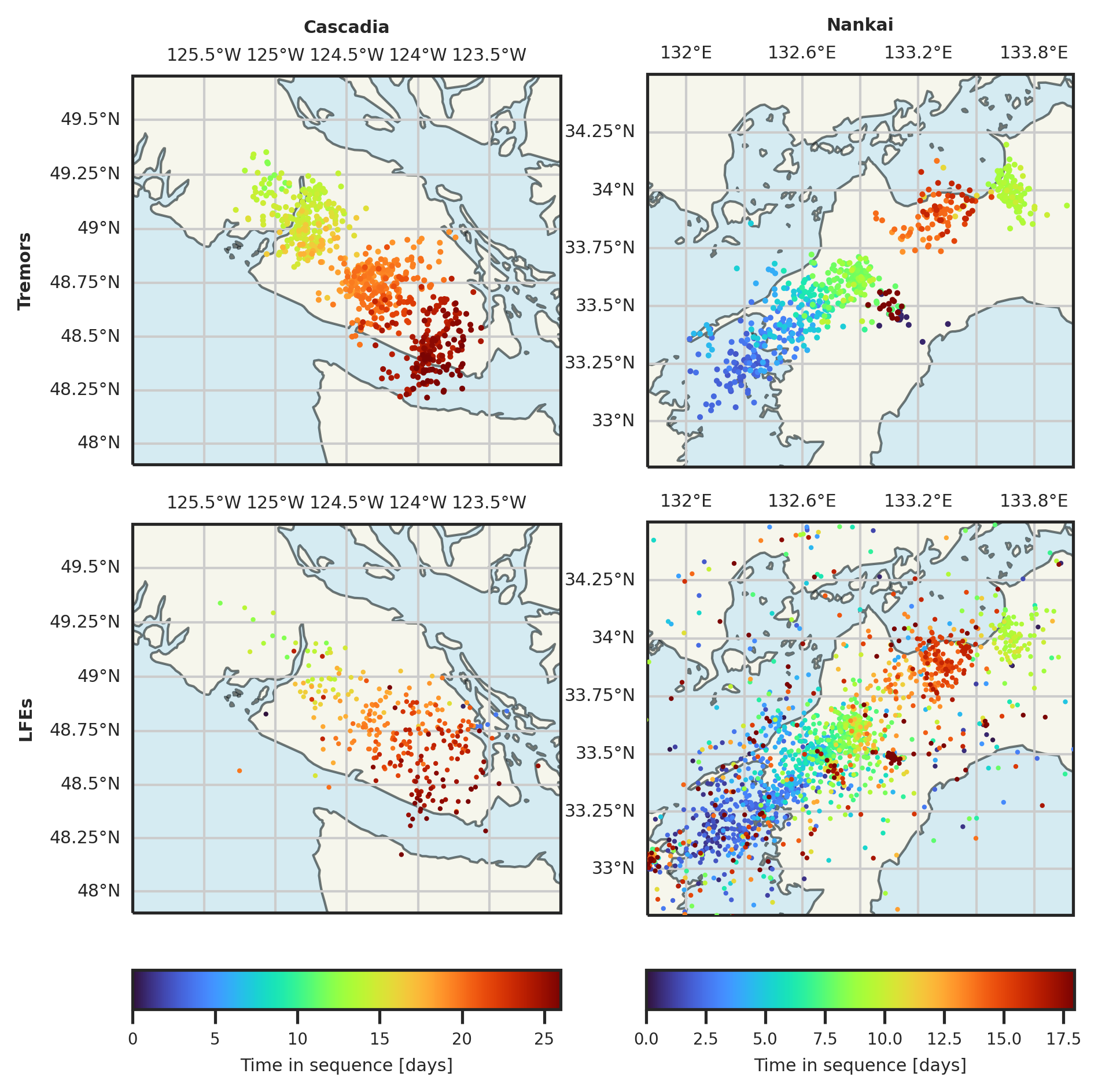}
\caption{Tremor and LFE detections in Cascadia and Nankai, colored according to the time within the sequence. Tremors detections have been filtered using DBScan \cite{wechCatalogingTectonicTremor2021}. For tremor locations, we add a small scatter (Gaussian with standard deviation 0.03\degree) to reduce overlap caused by the quantization of the grid search (4~km). Note that the LFE detections are identical to those presented in \cite{munchmeyerDeepLearningDetects2024}.}
\label{fig:reference_results}
\end{figure*}

We apply our tremor and LFE detection workflows to shorter time periods in Nankai and Cascadia (Figure~\ref{fig:reference_results}). 
The results for the reference regions stand in clear contrast to the findings in Chile.
For the tremor detection, a substantial number of detections remains even after the clustering, even though the clustering parameters are effectively more strict in the reference regions as we do not use overlapping cells.
The recovered clusters show a clear migration pattern, a typical feature of tremors that is interpreted as tracking slow slip on the interface \cite{michelSimilarScalingLaws2019}.
No such tremor migration patterns have been observed in Chile (except for the offshore active seismic experiments).
The LFE detections show migrations consistent with the tremor migrations in both Cascadia and Nankai, even thought we detected them with independent and orthogonal methods.
However, we note that not all regions hosting LFEs show such obvious migration patterns, e.g., migrations in Guerrero, Mexico, tend to be more subtle \cite{frankUsingSystematicallyCharacterized2014}.

As for Northern Chile, we perform template matching on the LFE detections to verify their LFE characteristics (Figure~\ref{fig:lfe_examples_reference}).
Afterwards, we conduct the same deduplication and classification scheme as before.
In both regions, several examples fulfill all criteria (P- and S-waves, no time-of-day bias, a clustered temporal distribution, and a location close to the plate interface) set out for LFEs.
We note that the total number of examples after the deduplication is substantially lower than the number of active LFE templates reported in regional catalogs during these periods \cite{bostockMagnitudesMomentdurationScaling2015,katoDetectionDeepLowfrequency2020}.
This confirms our assumption that the method is merging distinct LFE families into one example.
While this is not problematic for identifying whether a region hosts LFEs at all, for a complete LFE catalog the procedure needs to be repeated with more strict criteria for merging event families.

As for Chile, we locate examples with two distinct phase arrivals to verify their LFE characteristics.
For Cascadia, the epicenter of all located examples fall within the tremor band with depth ranging from 32 to 38~km.
This is consistent with the depth range for LFEs determined by \citeA{bostockMagnitudesMomentdurationScaling2015}.
For Nankai, we located a total of four examples.
Three of these locate within the known LFE areas at depth around 31~km, consistent with the LFE catalog of \citeA{katoDetectionDeepLowfrequency2020}.
The last family locates in close proximity to the Torigatayama limestone mine, a mine using blasting \cite{nittetsu2016torigatayama}, at a depth of only 140~m.
Our determined location is about 2.5~km from the mine, well within the inferred horizontal uncertainty of 6.2~km.
Similar to Chile, the mining events in Nankai produce a long coda and clear surface waves (Figure~\ref{fig:lfe_examples_reference} right column).
In contrast, the LFE traces have shorter S wave trains and no obvious surface waves (Figure~\ref{fig:lfe_examples_reference} left and center column).

\section{Discussion}

\subsection{Comparing regional detection sensitivity}
\label{sec:sensitivity}

\begin{figure*}[ht!]
\centering
\includegraphics[width=\textwidth]{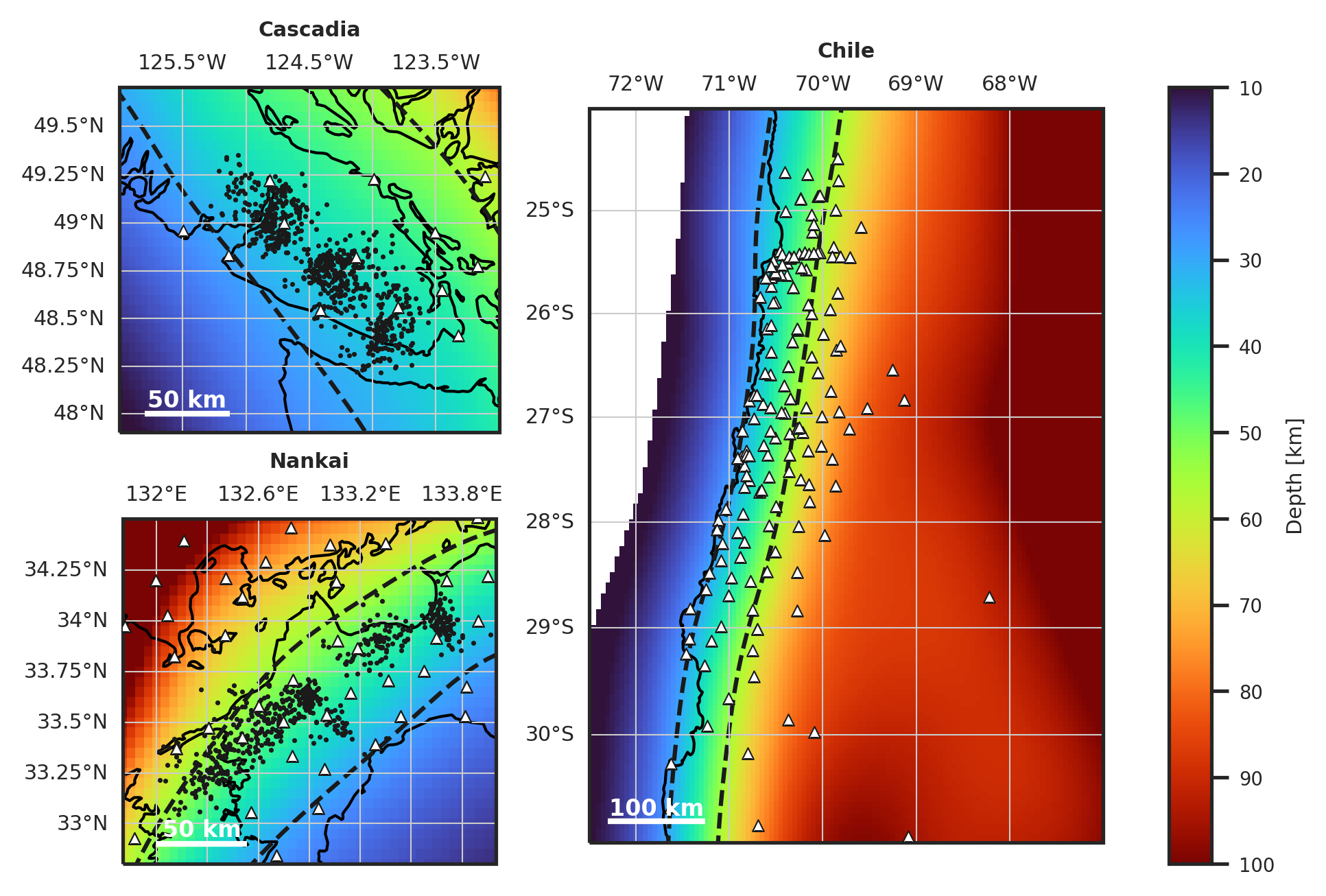}
\caption{Slab geometries \cite{hayesSlab2ComprehensiveSubduction2018a}, station coverage, and tremor detections in Cascadia, Nankai and Northern Chile. We indicate stations with white triangles. The median distance to the fifth closest station is 60~km in Cascadia, 37~km in Japan, and 27~km in Chile. For Cascadia and Nankai, we show tremors detected in this study using black dots. As in Figure~\ref{fig:reference_results}, we add a small scatter to the individual tremor locations. Black dashed lines show the 30~km and 55~km isodepth lines, a wide estimate of the typical interface depth of tremor-genic areas.}
\label{fig:slab_models}
\end{figure*}

Several factors determine the detection sensitivity of LFE and tremor detection workflows.
The two major factors are the network coverage and the SNR at each individual station.
The SNR in turn can be decomposed into the noise background (denominator) and the signal amplitude (numerator).
In the following, we will compare these two terms between the regions to estimate differences in detection sensitivity.
We note that we design our workflows to be independent of the amount of data or only have expected improvements in the sensitivity with longer duration, i.e., we ensure that the sensitivity due to the changes in study area and duration do not negatively impact the comparison.
For tremor detection, the methods is independent of the study duration, with the longest range dependencies coming from the clustering on a scale of hours.
For the LFE detections, only the template matching is dependent on the amount of data.
As longer study durations will lead to potentially more detections and better waveform stacks, no degradation from longer durations is expected.

Comparing the station densities, we observe the highest density in Chile, followed by Nankai and then Cascadia (Figure~\ref{fig:slab_models}).
As a quantitative measure, we estimate the median distance to the fifth closest station for each station in the network, given we require at least five stations for the envelope correlation.
This distance is 60~km in Cascadia, 37~km in Japan, and 27~km in Chile.
For Chile, we excluded the temporary stations during the offshore active seismic cruise, as these were only active for a short time frame.
Still, the estimated density is an optimistic estimate, given that not all networks were running throughout the whole study duration.
Nonetheless, the station coverage is at least on par with the coverage in Japan throughout the majority of the study period.

The signal amplitude at a station is determined by the combination of source effects and path effects, with the latter dominated by geometric spreading and attenuation along the source-receiver path.
A first order proxy for the effects of spreading and attenuation is the distance between stations and source regions, which depends on station coverage and slab geometry.
Figure~\ref{fig:slab_models} compares the regional slab geometries and station coverages between Northern Chile and the two reference regions.
In both Cascadia and Nankai, deep tremors occur within a band between roughly 30~km and 50~km interface depth.
The tremors in Nankai occur on average slightly further downdip than in Cascadia.
For the slab geometry in Chile, this depth band of the interface is almost fully underneath the land, i.e., in a region well covered with seismic stations.
Dense station coverage extends down to slab depth above 90~km, covering the full zone of the reported deep SSEs \cite{kleinReturnAtacamaDeep2022}.
As the 30~km depth contour of the plate interface coincides with the coastline, shallower tremors might be difficult to detect.
However, \citeA{munchmeyer2024chile_eqs} document that the downdip end of interface seismicity is consistently below land in this area.
As tremors typically occur in the transition zone downdip of the interface earthquakes, this provides another indicator that potential tremors in Northern Chile would be expected to happen in an area with good station coverage.
Overall, source to station distances should be similar in Chile compared to the reference regions, i.e., no substantial differences in path effects are to be expected, and therefore no substantial difference in expected amplitudes for a given source.

In contrast to deep tremors, shallow tremors occur at depth shallower than 20~km.
This area is fully offshore and therefore not covered by our network.
While the distance between shore and trench is comparably low in Chile, signals from shallow tremors would be attenuated substantially more strongly than those of deep tremors.
Detecting shallow tremor in other world regions required offshore instrumentation, such as the S-Net deployment in the Japan trench \cite{nishikawaSlowEarthquakeSpectrum2019}.

\begin{figure*}[ht!]
\centering
\includegraphics[width=\textwidth]{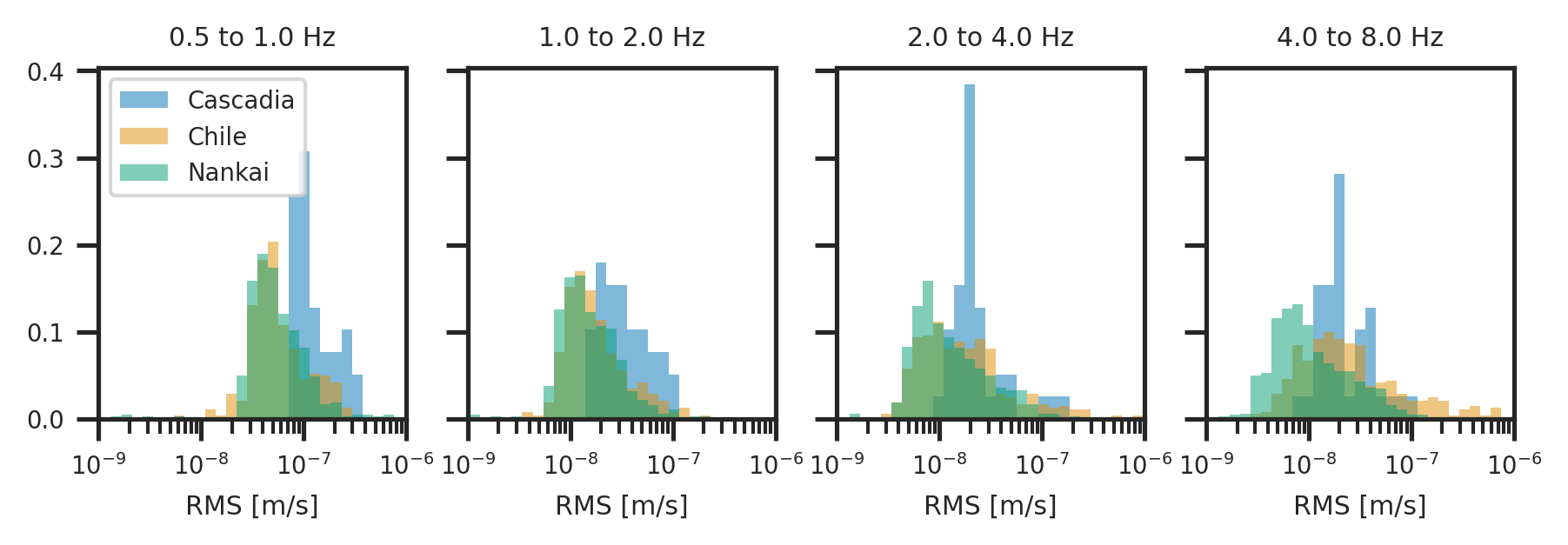}
\caption{Comparison of the background noise across the three regions. At low frequencies, the noise levels are similar in Chile and Nankai and higher in Cascadia. At higher frequencies, the noise is lowest in Nankai and at a similar, higher level for Cascadia and Chile. For each stations, we calculate the root mean square (RMS) of the signal in four distinct frequency bands in non-overlapping 60~s windows. We then select the 90th percentile among all time windows to get an estimate of the high noise condition at each station while not being affected by short transients. The histograms show the distributions of these percentile values among the different stations. We normalize the histograms using the station count for better visibility.}
\label{fig:region_noise}
\end{figure*}

As the denominator of the SNR, we evaluate the background noise in the typical tremor band (Figure~\ref{fig:region_noise}).
At frequencies between 0.5 and 1.0~Hz, Chile and Nankai show similar noise distributions, while the noise in Cascadia has roughly double the root mean squared (RMS).
At higher frequencies, the noise in Nankai is systematically lower than in Chile, with a median factor of 2.4 at frequencies between 4 and 8~Hz.
The lower noise level in Nankai is likely due to the borehole installations of the HiNet stations.
As large parts of the ambient noise field are composed of surface waves, the borehole stations are particularly useful at higher frequencies.
At higher frequencies, the noise levels in Chile approach the ones in Cascadia, with similar median noise levels between 4 and 8~Hz.
Due to the very different environmental conditions and seismic background rates, we suspect that the origins of these noise fields are different.
Nonetheless, they should have comparable impact on the detection sensitivity.
We note that a handful of stations in Chile show substantially higher noise than the median, while no similar outliers exist for Nankai or Cascadia.
This is likely because the location selection for the temporary stations in Chile was less rigorous than for the permanent deployments in Cascadia and Nankai.
However, due to their small number, these stations do not matter for the sensitivity analysis of the seismic network in its entirety.

In addition to SNR and station coverage, the actual shape of the waveform will be relevant for tremor and LFE detection.
The envelope correlation method employed for tremor search depends on the low-frequency coherence of the envelope across stations.
In our study, we low-pass filter our envelopes at 0.2~Hz, i.e., we only consider slow variations in the envelope.
For such variations to become incoherent across stations, a complex subsurface structure or strong site effects are required.
There is no evidence for either of the two in Chile, as shown by tomography results \cite{pasten-arayaAlongDipSegmentationSlip2022} and by low station residuals in magnitude scale calibration \cite{munchmeyer2024chile_eqs}.
We thus assume that envelope coherence across stations does not differ substantially between the studied regions.
The impact of waveform shape on LFE detection is substantially more difficult to analyze, due to the low interpretability of the underlying deep learning model.
However, two indicators suggest that the deep learning model is transferable across regions.
First, transferability for the LFE detection model was explicitly tested for four regions, including our two reference regions, in \cite{munchmeyerDeepLearningDetects2024}.
In particular, these results show that the model can correctly detect LFEs in regions without any examples from these regions required in training.
Second, phase pickers for earthquakes are transferable across regions, even though in-region performance is slighly superior to cross-region performance \cite{munchmeyerWhichPickerFits2022}.
We therefore suggest that in general the model should be able to identify LFEs in Chile, if the SNR conditions are similar.

Putting together the factors above, we can provide rough estimates for the size of tremors and LFEs that could exist and that might be detectable with our methods in Chile.
In comparison to Cascadia, we expect to be able to detect tremors of the same magnitude, or even slightly smaller.
This is due to the superior station density and slightly more favourable noise conditions in Chile.
In comparison to Nankai, we should be able to detect tremors with at least about twice the velocity amplitude.
Under similar spectral characteristics, this would roughly correspond to twice the moment rate.
We estimate this factor from the difference in noise levels in the least favorable band, i.e., 4 to 8~Hz.
However, this is likely a conservative estimate, given that we have a more dense station coverage and a lower difference in noise levels at lower frequencies, which might already be sufficient to detect tremors.
Typical moment rates for tremors, LFEs, and SSEs, are on the range of $10^{11}$ to $10^{13}$~Nm/s \cite{ideBridgingGapSeismically2008,aguiarMomentReleaseRate2009}.
Note that such moment rates for SSEs are geodetic long-term averages, which can be exceeded by several orders of magnitude on short time scales \cite{frankDailyMeasurementSlow2019}.
Consistent with our findings on sensitivity, typical observed moment rates are higher in Cascadia than in Nankai \cite{ideSlowEarthquakeScaling2023}.
This suggests that we should be able to detect tremors with moment rates in this range.

\subsection{The 20th September 2019 tremor reports}
\label{sec:pasten_araya}

While no large-scale tremor catalog exists for the region, \citeA{pasten-arayaAlongDipSegmentationSlip2022} report tremor detections around the subducted Copiapó ridge.
Analyzing 8 months of data, the authors identified nine tremor occurrences, all on 20th September 2019 between 00:39 and 03:52 (UTC) before the beginning of our study period.
For detecting the tremors, the authors used a very similar envelope correlation method to the one presented here, however, with slightly different preprocessing steps and a 1D velocity model.
In addition, all detections have been manually inspected to exclude examples with clear P and S arrivals.
\citeA{pasten-arayaAlongDipSegmentationSlip2022} infer tremor locations towards the downdip end of the interface seismicity band or within the transition zone.
The most likely depth estimates are slightly above the plate interface, even though there is substantial uncertainty.

Here, we reprocess the data from the study of \citeA{pasten-arayaAlongDipSegmentationSlip2022} for 20th September 2019 between 00:00 and 04:00 (UTC).
As before and as in \cite{pasten-arayaAlongDipSegmentationSlip2022}, we use the envelope correlation method by \citeA{wechCatalogingTectonicTremor2021} , but now with the preprocessing described in \cite{pasten-arayaAlongDipSegmentationSlip2022}.
However, for higher location accuracy we use the 3D velocity model from \citeA{munchmeyer2024chile_eqs}.
With this processing, we were able to reproduce six of the detections from \cite{pasten-arayaAlongDipSegmentationSlip2022}.
We visualise their waveform envelopes, locations and location uncertainties in Figures~\ref{fig:pasten_enveloc1} to \ref{fig:pasten_enveloc6}.
Out of the six detections, four have their most likely depth at 103~km, the deepest depth considered in our search.
The remaining two examples locate at 59~km depth, with the most likely depth range between 40 and 80~km, and at 43~km but with an uncertainty ellipse including almost the complete search range.
The epicenters show a wide scatter both onshore and offshore.

Given the characteristics of the reported signals and our reprocessing, it seems uncertain if the detections can be attributed to tectonic tremors.
First, our reprocessed locations are not compatible with a source on the plate interface.
The four deep locations suggests signals from outside the network or deeper below the network.
The remaining two detections are both still deeper than the plate interface, which is only included in the uncertainty ellipse of the event with the poorest depth constraint.
Second, that tremors would only occur within a four hour window of the full 8~months study period seems unlikely.
Typical tremor sequences in other regions consist of tremors over days to weeks \cite{behrWhatThereStructures2021}.
In addition, our study did not produce further detections in this region with a similarly dense and substantially wider seismic network spanning 3.5 years.
Based on this information, we consider that the detections from \citeA{pasten-arayaAlongDipSegmentationSlip2022} are likely not tectonic tremors, even if we are not able to identify the origin of these events.
While, for example, teleseismic sources might cause the signals, it remains unclear why these detections should only occur on a single day either.

\subsection{Limitations of our detection methods}

Our study has not found any evidence of either tectonic tremors or LFEs in Northern Chile between 24\degree S and 31\degree S.
However, this conclusion relies on several assumptions about the nature of LFEs and tremors that are important to revisit to explore the limitations of our study.
Starting with tremors, the first assumption is that the tremors can be detected with the envelope correlation method.
For this, the tremors need to be substantially above the noise in the frequency band from 1 to 8~Hz, and have a duration that can reasonably dominate a 300~s correlation window.
In addition, the tremor source process needs to be sufficiently non-smooth in time to provide robust differential times that can be identified through cross-correlation.
Second, to pass our clustering criteria, the tremors need to occur in bursts, because we would miss isolated tremors by design given that we filter detections with DBScan \cite{wechCatalogingTectonicTremor2021}.
Third, we assume that tectonic tremors occur on the plate interface or in its immediate vicinity.
Of these assumptions, probably the most questionable point is a sufficient SNR, as discussed in Section \ref{sec:sensitivity}.
It is therefore possible, that a study with, for example, borehole sensors or array processing on a more dense network, might be able detect tremors if they exist.

Similar to the tremor analysis, potential missed LFE detections might originate from both the initial detection step and the subsequent reprocessing.
As the initial detection step is based on a deep learning model, it is difficult to identify the signal characteristics required to detect potential LFEs.
However, \citeA{munchmeyerDeepLearningDetects2024} showed that the LFE detection model transfers successfully between Cascadia, Guerrero, and Nankai and similar transferability results exist for regular earthquake detection models \cite{munchmeyerWhichPickerFits2022}.
This suggests a certain universality of LFE and earthquake signal shapes across regions.
Therefore, it seems unlikely that LFE signals in Chile would have a sufficiently different signal shape to be missed by the detection model.
However, as for tremors, a worse SNR might lead to missed detections, if LFEs have smaller than expected magnitudes.

For the template matching of LFE candidates, we assume that LFEs are repeating, with a requirement of at least 100 detections of the same family.
This bar is low, compared to the observed repeat rates in other regions, but we can not rule out the existence of LFEs with low-repeat rate \cite{katoDetectionDeepLowfrequency2020,shelly15YearCatalog2017,frankUsingSystematicallyCharacterized2014,bostockMagnitudesMomentdurationScaling2015}.
The other factor that might cause low detection counts are templates of insufficient quality.
This could, again, occur due to unfavorable signal-to-noise conditions.
However, it is worth pointing out that at the signal-to-noise conditions in the reference regions, the templates were of sufficient quality.

\subsection{Implications of the missing tremor and LFE detections}

As our analysis did not find any evidence of tremors or LFEs in Northern Chile, we will now discuss the potential reasons and implications of this null result.
We propose four potential explanations: (i) tremors/LFEs have smaller moment rates than in other regions; (ii) the region hosts tremors/LFEs but none happened during our study period; (iii) the region does not host any tremors/LFEs; (iv) tremors/LFEs have different characteristics in this region than in others.

For option (i), it is possible that we are not able to detect LFEs and tremors because they would be hidden below the noise.
\citeA{frankDailyMeasurementSlow2019} show that LFE moment rate scales with the rate of geodetic slip and demonstrate its variations reflect the intermittence of slow deformation \cite{jolivetTransientIntermittentNature2020}.
Similarly, \citeA{ideSlowEarthquakeScaling2023} suggest that SSEs, tremors and SSEs, while at different scales, follow a consistent scaling with a region-dependent upper bound on their moment rate.
As we showed in our analysis of the station geometries and noise conditions, we expect to see deep tremors with moment rates above $10^{12}$~Nm/s to $10^{13}$~Nm/s.
Yet the SSEs in the region only have a moment rate between  $6 * 10^{11}$~Nm/s (2014 deep SSE \cite{kleinDeepTransientSlow2018}) and $3 * 10^{12}$~Nm (2023 shallow SSE \cite{munchmeyer2024chile_sse}), close to the lower boundary of our detection capabilities.
While these are only two positive observations, it is noteworthy that continuous GNSS observations have not identified SSEs with higher moment rate, suggesting that such events are, at least, rare.
In addition, it is unclear to which degree the slip during the observed SSEs is continuous or intermittent.
It is therefore possible, that tremors and LFEs might have moment rates below our detectability threshold.
For shallow tremors, this scenario is more likely, as no offshore instrumentation was employed, lowering the detection sensitivity offshore.

To improve the detection threshold, either better instrumentation or more dense instrumentation with different methods would be required.
Borehole instruments could reduce the ambient noise, thereby improving SNR.
Dense instrumentation, such as seismic (mini-)arrays, would enable the application of beamforming methods that can use the coherency of high-frequency signals to distinguish waves from different origins.
However, due to the combinatorial explosion when considering sources at high frequencies, such methods are still computationally challenging.
For shallow tremor, that would occur offshore, ocean bottom instrumentation might be required.
For example, the S-Net ocean bottom deployment in the Japan trench enabled the detection and characterisation of shallow tremors there \cite{nishikawaSlowEarthquakeSpectrum2019}.

Option (ii), the possibility that LFEs and tremors occur in Chile but not during the 3.5 year time period analyzed here, is supported by the timing of deep SSEs in the catalog.
Deep SSEs have been observed in 2014 and 2020, with further possible detections in 2005 and 2009 \cite{kleinReturnAtacamaDeep2022}.
Our 3.5~year study period falls into the gap between 2020 and the next expected recurrence around 2025.
However, a 3.5 year quiescence of tremors and LFEs in a region generally hosting such events would be unique among the subduction zones with known tremor activity.
Well studied regions, like Nankai, Guerrero, or Cascadia, show numerous tremor and LFE detections in between the larger burst associated with detected SSEs \cite{frankSlowSlipHidden2016,michelSimilarScalingLaws2019,katoDetectionDeepLowfrequency2020}.
On the other hand, modeling studies suggest that complete tremor/LFE quiescence between SSEs is possible with essentially unbounded return periods \cite{wangPulselikeRupturesSeismic2023}.

This leads us to option (iii), that the region is not hosting LFEs and tremors at all.
While most regions hosting SSEs also have records of tremors or LFEs, some examples, do not have a comprehensive record of tremors or LFEs associated with them \cite{montgomery-brownTremorgenicSlowSlip2015,mouchonSubdailySlowFault2023}.
Several possible causes have been proposed, including fluid availability, thermal properties, pressure, and temperature \cite{montgomery-brownTremorgenicSlowSlip2015}.
Here we discuss some of these properties with respect to Northern Chile.

LFE and tremor occurrences have been linked to fluid rich environments detected through geophysical imaging \cite{katoVariationsFluidPressure2010,delphFluidControlsHeterogeneous2018,behrWhatThereStructures2021}.
In Northern Chile, one main source of fluids, sediments entering into the interface at the trench, is very weak.
Due to the very arid climate in Northern Chile only low amounts of sediments are available at the trench, with a sediment layer of only about 200~m thickness \cite{vankekenSubductionFactoryDepthdependent2011}.
Nonetheless, plenty of other indicators highlight that fluids are ubiquitous in the Northern Chilean subduction.
The region hosts recurrent swarm activity, typically linked to fluid migrations \cite{holtkampEarthquakeSwarmsSouth2011,marsanEarthquakeSwarmsChilean2023,ojedaSeismicAseismicSlip2023,munchmeyer2024chile_sse}, shallow SSEs, linked to reduced normal stress from fluids \cite{munchmeyer2024chile_sse,frankAlongfaultPorepressureEvolution2015,warren-smithEpisodicStressFluid2019,fargeEpisodicityMigrationLow2021,danrePrevalenceAseismicSlip2022,bangsSlowSlipHikurangi2023}, strong intraplate activity in the subducting plate, linked to dehydration in the blueschist to eclogite transition, and mantle wedge seismicity, caused by fluid intrusion into the cold mantle wedge \cite{munchmeyer2024chile_eqs}.
These fluids can enter the subducting plates due to fractures from the numerous outer rise earthquakes observed in the region \cite{munchmeyer2024chile_eqs}.
Additionally, in the Copiapó ridge, seamounts enter the subduction, which have repeatedly been linked to fluid incursions through both sediment transport and plate fractures \cite{shaddoxSubductedSeamountDiverts2019,gaseSubductingVolcaniclasticrichUpper2023,bangsSlowSlipHikurangi2023,chesleyFluidrichSubductingTopography2021,sunMechanicalHydrologicalEffects2020}.
Seismic imaging studies found elevated vp/vs ratios in the potentially tremorgenic zones around the Copiapó ridge, suggestive of fluids on the interface \cite{pasten-arayaAlongDipSegmentationSlip2022}.
However, even without fluids it is possible to model SSEs through geometric complexities \cite{romanetFastSlowSlip2018} or thermally activated friction \cite{wangPulselikeRupturesSeismic2023}.

The role of the temperature of a subduction zone for SSE, tremor, and LFE occurrence is debated \cite{montgomery-brownTremorgenicSlowSlip2015}.
Most tremor detections stem from hot subduction zones, such as Nankai or Cascadia.
\citeA{abersColdRelativelyDry2017} relate the thermal state of a subduction zone to the hydration of the mantle wedge, suggesting that hotter subduction zones have a higher abundance of fluids.
As tremors and LFEs have been linked to reduced normal stress due to fluid pressure, this could explain why these regions might have favorable conditions for tremors and LFEs.
However, some cold subduction zones host tremor and LFEs, for example, Hikurangi \cite{aden-antoniowLowFrequencyEarthquakesDowndip2024}.
Therefore, it is unclear what influence the temperature conditions in northern Chile might have on the observed absence of tremors and LFEs.

While aseismic slip rates might lead to tremors/LFEs with moment rates below our detection sensitivity, as discussed above, they might also cause a complete absence of tremors and LFEs.
The link between tremor/LFE signals and SSEs is complex, with tremors appearing in parallel to aseismic deformation \cite{michelSimilarScalingLaws2019}, tremors fronts following aseismic deformation \cite{bleteryCharacteristicsSecondarySlip2017}, and temporary quiescence of tremors during aseismic deformation \cite{wechSlipRateTremor2014}.
Several studies proposed that too low \cite{wechSlipRateTremor2014,frankUsingSystematicallyCharacterized2014,frankDailyMeasurementSlow2019} or too high \cite{montgomery-brownTremorgenicSlowSlip2015} slip rates might lead to a lack of tremors and LFEs.
Both the deep and shallow SSEs around the Copiapó ridge have slip rates above $10^{-6}$~m/s \cite{kleinDeepTransientSlow2018,munchmeyer2024chile_sse}, the boundary value estimated by \citeA{montgomery-brownTremorgenicSlowSlip2015}.
However, this geodetic estimate is a long-term average, not resolving potential intermittence of slip rates that might be essential for tremor- and LFE-genesis \cite{frankDailyMeasurementSlow2019}.
The smoothness/roughness of the slow deformation process in Chile might be controlling the tremor/LFE production.

Looking at the occurrence of LFEs and tremors, it is worth not only considering macroscopic features but also looking at the microstructure.
Fine-scale tremor and LFE maps show that the events do not occur on the full interface, but rather in distinct and long-term stable patches with diameters of only few kilometers \cite{shellyPreciseLocationSan2009,armbrusterAccurateTremorLocations2014}.
This suggests structural heterogeneities on the interface, with only certain patches fulfilling the structural requirements of tremorgenesis.
\citeA{mouchonSubdailySlowFault2023} linked this segmentation to the non-linearity in scaling between SSE and LFE activity in Guerrero, Mexico, showing that parts of the interface accommodating slow slip might not be able to accommodate LFEs.
While the exact structural properties of these patches are not fully understood \cite{behrWhatThereStructures2021}, the observed segmentation suggests that it is possible that a subduction zone does not host the structures required for LFEs and tremors at all.
This would suggest that the northern Chilean subduction zone could host SSEs that are not accompanied by tremor or LFEs.

Lastly, option (iv), the possibilities that LFE/tremor have fundamentally different characteristics in Northern Chile, is essentially a wild card.
As discussed in the previous section, our detection algorithms rely on assumptions about the repeat rate, clustering, spectra and signal shapes of the LFEs/tremors.
While our comparison study shows that these assumptions are valid and justified in reference regions, such as Nankai and Cascadia, and consistent with the majority of models proposed for tremors and LFEs, it is possible that they are not valid in Northern Chile.
We address this problem by applying two orthogonal methods targeting both tremors and LFEs.
Nonetheless, we are unable to rule out the existence of tremors or LFEs with substantially different characteristics in Northern Chile.

\section{Conclusion}

In this study, we performed a systematic search for tectonic tremors and LFEs in the Atacama segment (24\degree S to 31\degree S) in Northern Chile.
We search for both tremors and LFEs with complementary methods to account for the different characteristics of these two signal classes.
The detection methods have previously been applied successfully to other study regions.
Despite the documented SSE cases, we did not detect any compelling evidence for tremor or LFE activity in the region.
While there is a small number of supposed detections for either tremors or LFEs for which we are not able to identify the source with certainty, their characteristics make a tectonic source on the interface unlikely.
In contrast, the identical workflows applied to Cascadia and Nankai successfully recover both LFE and tremor activity, together with their characteristic migration patterns.
We note that due to our network geometry, located exclusively onshore, our results focus on deep tremor and LFEs, while shallow events might be missed due to their higher distance to the network.

While our results can not fully rule out tremors or LFEs in the region, they provide bounds on their possible moment rates and recurrence times.
These results are an important constraint for modeling studies, for example, studying the impact of temperature and fluid availability on the existence of tremor and LFEs.
In addition, they serve as a basis for developing and testing scaling relationships for SSEs, tremors, and LFEs.
Lastly, our results provide guidance for which instrumentation and methods might be required to identify potential lower magnitude tremors in Chile and beyond.

\section*{Open Research}
For Chile, we use seismic data from the C, C1 \cite{fdsn_c1}, CX \cite{fdsn_cx}, IU \cite{fdsn_iu}, GE \cite{fdsn_ge}, RI, Y6 \cite{fdsn_y6}, XZ \cite{fdsn_xz}, 2V \cite{fdsn_2v}, 3V \cite{fdsn_3v}, and 9C \cite{fdsn_9c} networks.
For Cascadia, we use seismic data from the CN network \cite{fdsn_cn}.
The HiNet data for Japan is available from the NIED at \url{https://www.hinet.bosai.go.jp/}.
The envelope correlation code from \citeA{wechCatalogingTectonicTremor2021} is available at \url{https://github.com/awech/enveloc}.

\acknowledgments
This work has been partially funded by the European Union under the grant agreement n°101104996 (“DECODE”) and the ERC CoG 865963 DEEP-trigger. Views and opinions expressed are however those of the authors only and do not necessarily reflect those of the European Union or REA. Neither the European Union nor the granting authority can be held responsible for them.
We thank INPRES for making the data from the RI network available to us.
We thank INPRES and CIGEOBIO (Universidad Nacional de San Juan) for the data from the stations DOCA and PEDE.
We thank everyone involved in the deployment of the seismic and geodetic networks and the data management. 
The computations presented in this paper were performed using the GRICAD infrastructure (\url{https://gricad.univ-grenoble-alpes.fr}), which is supported by Grenoble research communities.

\clearpage
\appendix

\section{Tremor and LFE detection details}
\label{sec:method_detail}

\subsection{Tremor detection}

We detect and locate tremors using the envelope correlation method of \citeA{wechCatalogingTectonicTremor2021}.
The method correlates envelopes of vertical seismograms between multiple stations.
These envelopes are a proxy for the energy rates recorded at each station over time.
For a tremor, we expect a similar development of the energy rate with a moveout consistent with seismic wave propagation.
The method by \citeA{wechCatalogingTectonicTremor2021} identifies the number of station pairs with correlation values above a threshold (here 0.7).
For each of these pairs, a differential travel time is identified by maximizing the correlation value.
If at least 5 station pairs exceed the correlation threshold, a grid search is used to locate the source of the coherent signal.
For the grid search, the method assumes an S wave moveout speed.

To apply the envelope correlation method to tremor detection, we follow the preprocessing of \citeA{wechCatalogingTectonicTremor2021}.
First, we bandpass filter the data between 1 and 8~Hz.
Second, we calculate the signal envelope, lowpass filter it at 0.2~Hz and downsample the envelope to 2~Hz.
For grid search, we use the 3D velocity model from \citeA{munchmeyer2024chile_eqs} for Chile and 1D velocity models for Cascadia and Nankai.
In all regions we use a grid spacing of 4~km in all axes and a search depth up to 100~km, well below the typical depth range of tectonic tremors.
We apply the tremor search to 5~minute windows with an overlap of 2.5~minutes between consecutive windows.

Due to the size of the study area in Chile, we split the region by latitude into bands of 1\degree width.
We overlap the bands by 0.5\degree, implying that each potential source is contained in two bands.
We process each band individually, using all stations within the band.
Due to the generous overlap between neighboring bands, this should not lead to a notable reduction in detection sensitivity.
Similar subdivisions are not necessary for the Cascadia or Nankai regions due to their smaller size.

While we aim to identify tremors, the envelope correlation method will detect any signal that leads to coherent envelopes across multiple stations.
In a seismically active region like Northern Chile, the method will identify earthquakes of sufficient size.
To remove earthquake detections, we compare the tremor candidates to a dense earthquake catalog \cite{munchmeyer2024chile_eqs}.
We remove all detections that have an earthquake at an epicentral distance less than 100~km within the detection window or within the minute preceding it.
We use this large distance threshold to account for the low location precision of the envelope correlation, in particular, at the boundary of the study region.
Nonetheless, even with this criterion it is statistically unlikely to exclude a detection based on an unrelated earthquake.

To further reduce the number of false detections, we cluster the detections using DBScan, keeping only events occurring within clusters \citeA{wechCatalogingTectonicTremor2021}.
This method uses the burst-like recurrence of tremors
We cluster the events based on their epicenter and the origin time, equating 1~h to 20~km.
We do not use the depth for clustering as it is not well-constrained.
We use DBScan with an epsilon of 20~km and a minimum number of 5 samples for a cluster.
We only retain detections within the clusters and discard isolated detections.
For Chile, we perform clustering on the concatenated detections from all subregions.
As this might lead to duplicate detections, the clustering parameters are effectively less strict in Chile than in the other regions.

\subsection{LFE detection}

Our LFE detection workflow builds upon the deep learning LFE detector from \citeA{munchmeyerDeepLearningDetects2024}.
The picker identifies P and S arrivals of LFEs at regional distances using a convolutional neural network architecture.
The model has been trained on combined data from Cascadia (Canada/USA), Nankai (Japan), Guerrero (Mexico), and the San Andreas fault (USA).
\citeA{munchmeyerDeepLearningDetects2024} have shown that the model is able to discover LFEs both inside the training regions and when applied to regions not included in training.
This suggests that the model might also be able to identify potential LFE signals in Northern Chile.

Similar to \citeA{munchmeyerDeepLearningDetects2024}, we integrate the picker into a detection workflow modeled after traditional earthquake detection.
First, we process the continuous data with the deep learning model to obtain potential P and S picks using SeisBench \cite{woollamSeisBenchToolboxMachine2022}.
We use a picking probability threshold of 0.1/0.15/0.15 (Cascadia/Chile/Nankai) for both P and S waves and an overlap of 50\% of the input window between subsequent applications of the model.
The different thresholds are motivated by the expierience from the original model development \cite{munchmeyerDeepLearningDetects2024}.
Second, we associate the picks using the PyOcto associator \cite{munchmeyerPyOctoHighthroughputSeismic2024}.
We require at least 13/10/12 total phase picks per event and 2/3/4 stations with both P and S pick.
We use region-dependent 1D velocity models for association (\citeA{bostockMagnitudesMomentdurationScaling2015} for Cascadia, \citeA{munchmeyer2024chile_eqs} for Chile, JMA2001 for Japan).
Third, we locate the LFE candidates using NonLinLoc \cite{lomaxProbabilisticEarthquakeLocation2000}.
For location, we use the 1D velocity models for Cascadia and Nankai and the 3D velocity model from \citeA{munchmeyer2024chile_eqs} for Chile.
We remove events with onset time uncertainties above 0.85/0.75/1.0~s.

Individual LFE waveforms are usually difficult to analyse visually due to their low SNR.
In particular, it can be difficult to decide whether an individual detection constitutes an LFE or not.
To alleviate this problem, we exploit the repetitiveness of LFE sources \cite{frankUsingSystematicallyCharacterized2014,shelly15YearCatalog2017}.
To this end, we perform template matching on our detections using the fast matched filter code \cite{beauceFastMatchedFilter2017}.
Due to the size of our study region, a template matching on all cells jointly would not be effective, as only the stations closest to the LFE source are expected to record its signals.
Instead, we subdivide the region into 19 overlapping cells of 1\degree ~by 1\degree (Figure~\ref{fig:lfe_cells}).
To account for changing station coverage we furthermore split our study into 7 time blocks (splitting on 2020-12-01, 2021-03-15, 2021-08-01, 2022-03-01, 2022-12-15, 2023-02-15, 2023-06-15, 2024-03-01).
See Figure~\ref{fig:chile_lfes_timing} for the stations counts per period.

For each space-time cell, we create a set of templates.
First, we identify the most promising set of stations by identifing the stations with the highest number of phase arrivals for events within this cell.
Note that the stations may locate outside the cell, which occurs in particular for offshore cells.
We select up to 10 stations with the highest phase count, excluding stations with less than 10 phase arrivals in total.
For each event within the spatial cell (even if outside the time range of the cell), we export a template.
Each template contains three component waveforms from the selected stations, bandpass-filtered between 1 and 8~Hz and resampled to 20~Hz.
The templates start 3~s before the S arrival and end 5~s after.
We use the picked S arrivals and supplement them with predicted travel times in case of missing picks.
We exclude events that are lacking waveforms at more than 2 stations.
We perform the same template export for Cascadia and Nankai, however, due to the size and duration of the reference analysis, we only use a single time block and one spatial cell (Cascadia)/two spatial cells (Nankai).

We perform template matching individually for each cell.
We match all events in the cell against continuous waveforms for the duration of the cell.
We normalize the correlation curve with the median absolute deviation (MAD) within 4~h windows.
All correlation values exceeding 8 * MAD are treated as detections, with a minimum separation of 5~s between two detections of the same template.
For each cell, we obtain a list of detections with the matching template, the time, and the normalized correlation value.
Furthermore, for each template we calculate a waveform stack by normalizing and summing up all its detections.

The output of the template matching allows for more elaborate analysis than the original detections.
First, the stacked waveforms of a repetitive event have better SNR, making both visual and automatic inspection of the templates more feasible.
Second, we can now study the group behaviour of the detections, such as how the rate of repeats or the change of activity level with time of day.
However, as the total number of detections still makes a fully manual inspection infeasible, we further group and filter the templates.
As LFEs are highly repetitive, it is likely that many templates identified the same families and will have highly similar matches.
To reduce work in manual inspections, we merge such templates.
To this end, we compute detection counts in non-overlapping 1~h bins and calculate cross-correlations between the detection rate of each template within a cell.
We merge all families with correlation values above 0.6, obtaining clusters of families.
For each cluster, we only analyze the template with the highest number of detections as this template is likely to be the most expressive,provide the best signal-to-noise ratio in the waveform stacks, and the clearest recurrence patterns.
We note that this approach might merge different LFE families as different families often show similar activity patterns.
However, as the goal of our study is not to obtain comprehensive LFE catalogs but to identify whether a region hosts LFEs at all, this more aggressive deduplication is acceptable.
To further reduce the number of clusters, we only keep clusters that contain at least two templates and at least one template with more than 100 detections.
This simplification relies on the assumption that LFE families have high numbers of repeats, making it highly unlikely that the original catalog found exactly one detection within a family.

For each cell, we manually inspect the remaining clusters.
For each cluster, we focus on three diagnostic plots describing the template with the most detections: the 3 component waveform stacks at all stations, a recurrence time plot throughout the time range of the cell, and a histogram of activity by day.
We classify the clusters based on four features (Figure~\ref{fig:lfe_classification}).
First, we identify if the time of day histogram shows a clear anthropogenic signature.
For example, many families show the characteristic timing of mine blasts or other mining related activities, such as elevated rates during the day with particular spikes in the early and late afternoon.
Second, we classify whether the event has a surface-wave-like waveform, with a narrow band of frequency content and with a smooth envelope.
Third, we count the number of clearly identifiable phase arrivals in the waveform stacks.
We do not require that this number of arrivals is visible at all stations, but label examples as containing two arrivals if they are visible at least at one station.
Fourth, we classify the recurrence pattern into three classes: LFE type, dense bursts, and clear separation between detections; bimodal distributions of detection times; other such as Poissonian recurrence typical for earthquakes.
For an LFE family, we would expect two phase arrivals with an LFE type recurrence and without an anthropogenic or surface wave signature.
For any feature, if in doubt, we assign the label expected for an LFE to include the family in later, more thorough, manual inspections.
We provide visual examples of the different feature manifestations in Figure~\ref{fig:lfe_classification}.

\bibliography{mybibfile,zotero}

\clearpage

\section*{Supplementary figures}
\FloatBarrier
\renewcommand\thefigure{S\arabic{figure}}
\renewcommand\thetable{S\arabic{table}}
\setcounter{figure}{0}
\setcounter{table}{0}

\begin{figure*}[ht!]
\centering
\includegraphics[width=\textwidth]{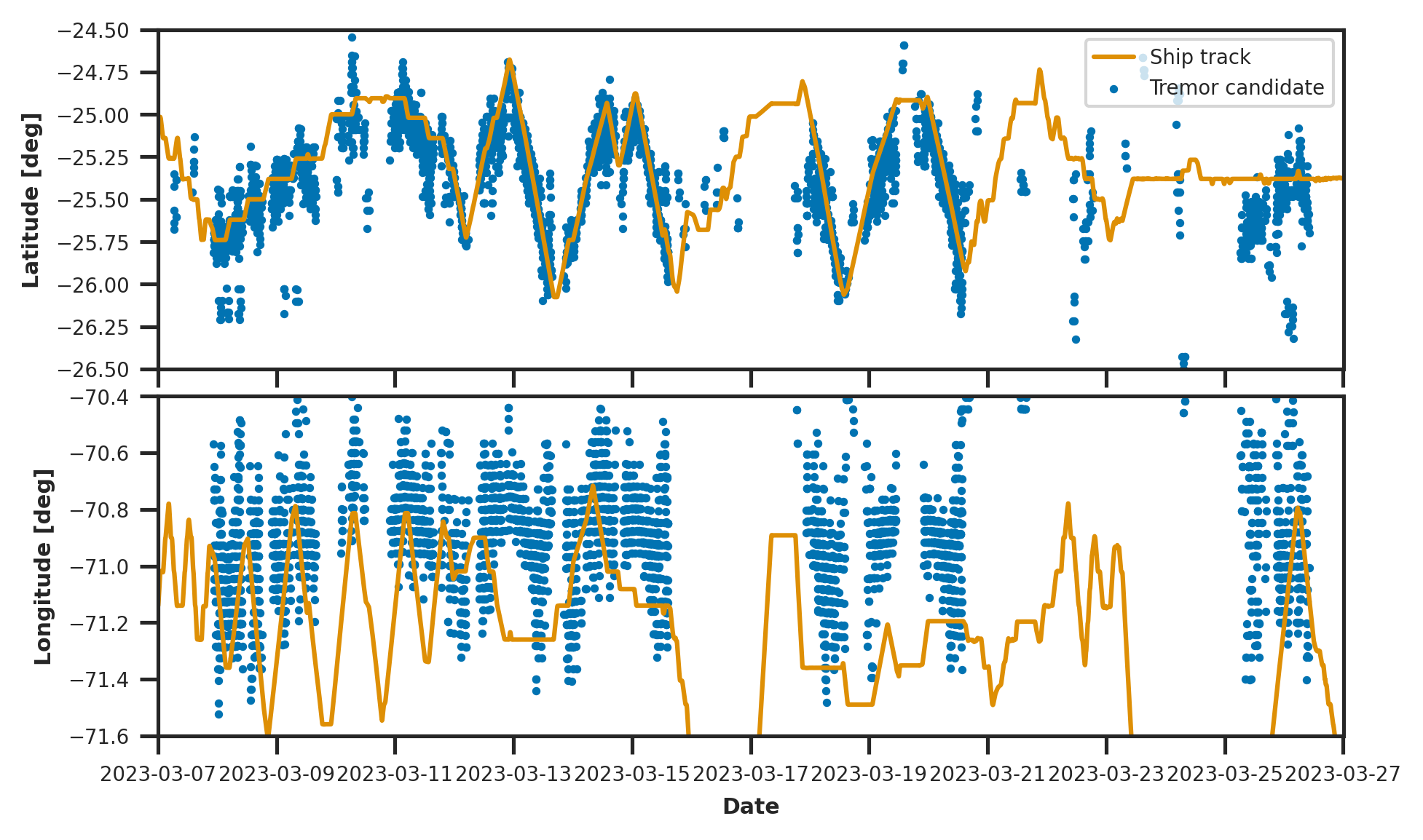}
\caption{Candidate tremor detections (blue) associated with the cruise SO297 of the RV Sonne in March 2023. The ship track is shown in orange. The top plot shows the latitude of detections and ship over time, the bottom plot the longitude.}
\label{fig:cruise_tremors}
\end{figure*}

\begin{figure*}[ht!]
\centering
\includegraphics[width=\textwidth]{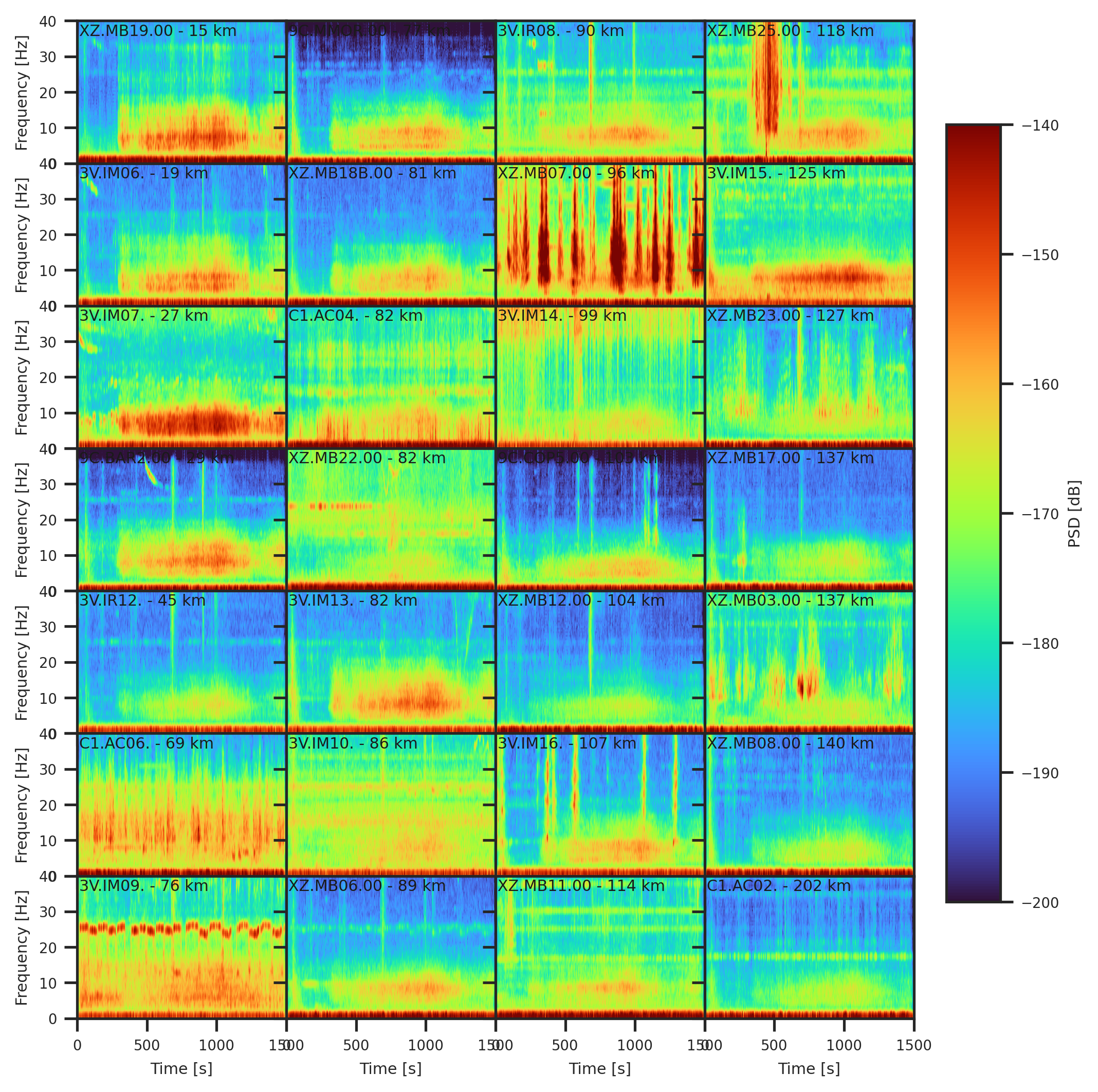}
\caption{Emergent 20~min signal on 2023-06-23, likely related to a marine or seafloor process. Each panel shows the power spectral density (PSD) for the vertical component waveforms of a station between 26.5\degree S and 28.5\degree S for the time from 2023-06-23 21:22:00 UTC to 2023-06-23 21:47:00 UTC. All stations are scaled equally. Stations are ordered by their distance to the most likely source of the 20~minute signal estimated using envelope correlation (Figure~\ref{fig:472_enveloc}).}
\label{fig:472_spectrogram}
\end{figure*}

\begin{figure*}[ht!]
\centering
\includegraphics[width=\textwidth]{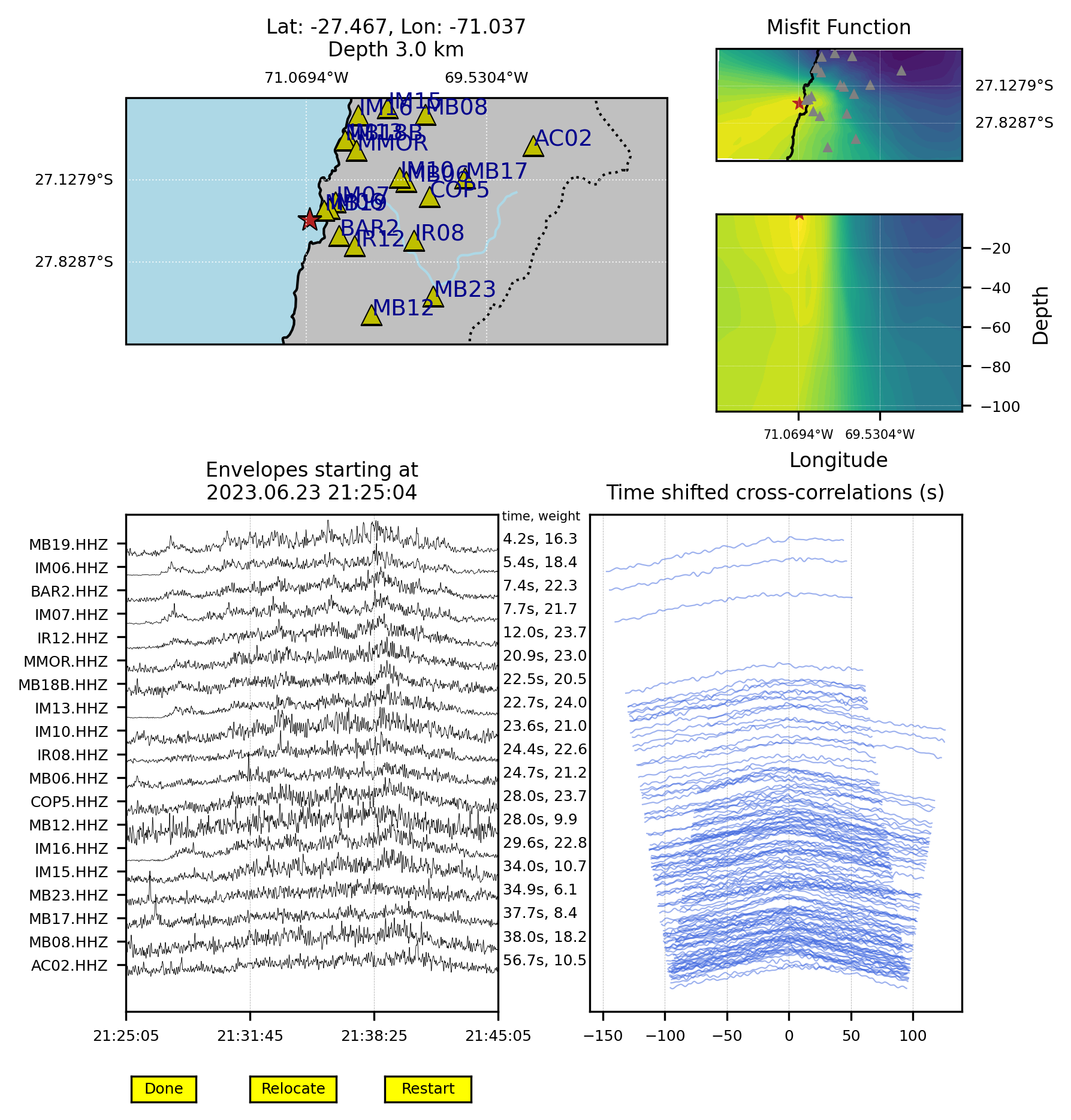}
\caption{Envelope correlation result for the emergent 20~minute signal around 2023-06-23 21:35 UTC. The most likely location for the source is at the seafloor, suggesting that the signal originates from a marine or seafloor process. The figure is based on the work of \citeA{wechAutomatedDetectionLocation2008}.}
\label{fig:472_enveloc}
\end{figure*}

\begin{figure*}[ht!]
\centering
\includegraphics[width=\textwidth]{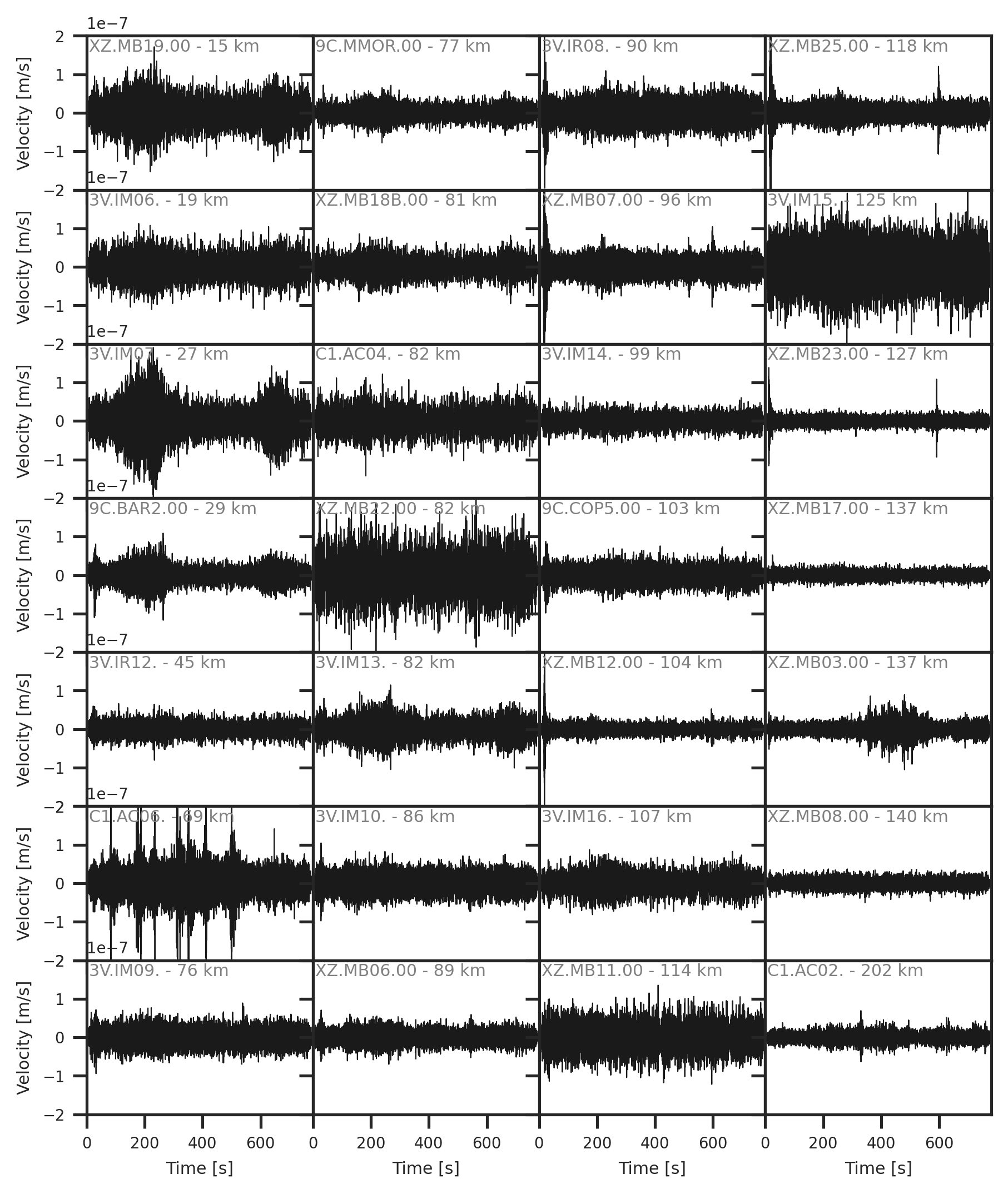}
\caption{Emergent signal on 2023-06-24, likely related to a marine or seafloor process. Vertical component waveforms for stations between 26.5\degree S and 28.5\degree S for the time from 2023-06-24 04:12:00 UTC to 2023-06-24 04:25:00 UTC. We restituted the instrument response. The data is bandpass filtered between 1 and 8~Hz. All stations are scaled equally. Stations are ordered by their epicentral distance to the most likely source of the 20~minute signal on the preceding day estimated using envelope correlation (Figure~\ref{fig:472_enveloc}).}
\label{fig:477_waveform}
\end{figure*}

\begin{figure*}[ht!]
\centering
\includegraphics[width=\textwidth]{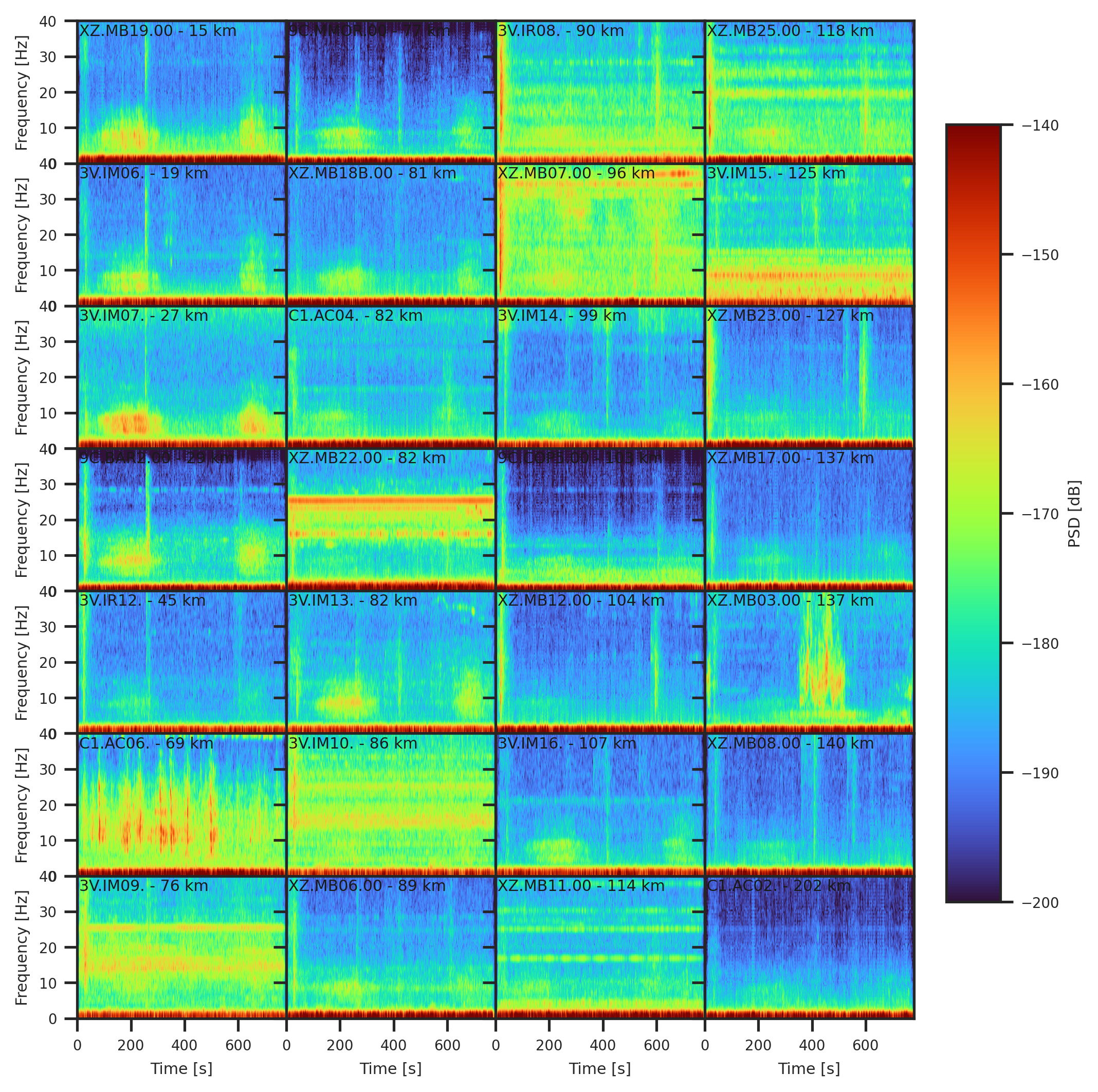}
\caption{Emergent signal on 2023-06-24, likely related to a marine or seafloor process. Power spectral density (PSD) for the vertical component waveforms for stations between 26.5\degree S and 28.5\degree S for the time from 2023-06-24 04:12:00 UTC to 2023-06-24 04:25:00 UTC. All stations are scaled equally. Stations are ordered by their distance to the most likely source of the 20~minute signal on the preceding day estimated using envelope correlation (Figure~\ref{fig:472_enveloc}).}
\label{fig:477_spectrogram}
\end{figure*}

\begin{figure*}[ht!]
\centering
\includegraphics[width=\textwidth]{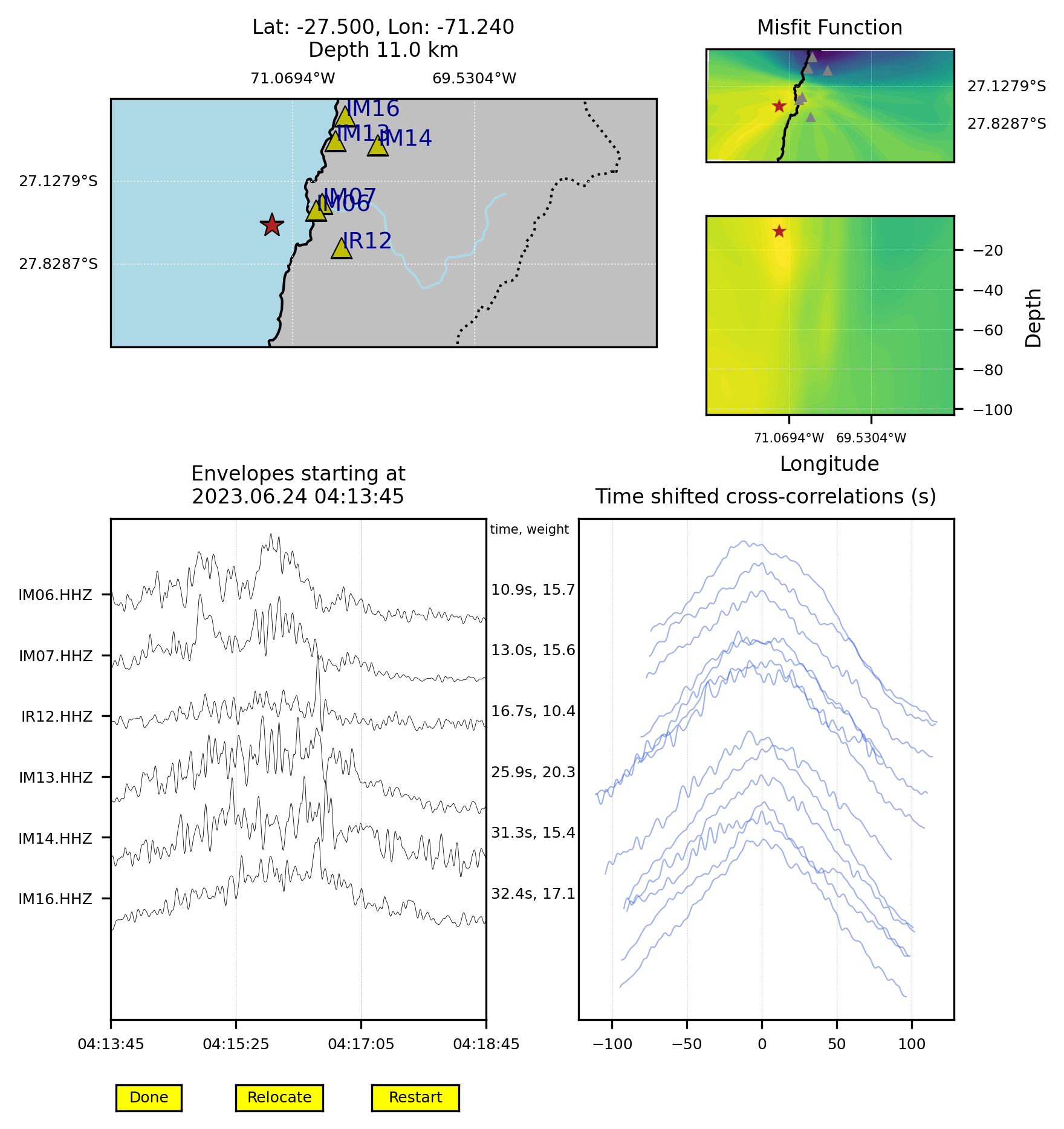}
\caption{Envelope correlation result for an emergent signal around 2023-06-24 04:16:00 UTC, likely related to a marine or seafloor process. The figure is based on the work of \citeA{wechAutomatedDetectionLocation2008}.}
\label{fig:477_enveloc}
\end{figure*}

\begin{figure*}[ht!]
\centering
\includegraphics[width=\textwidth]{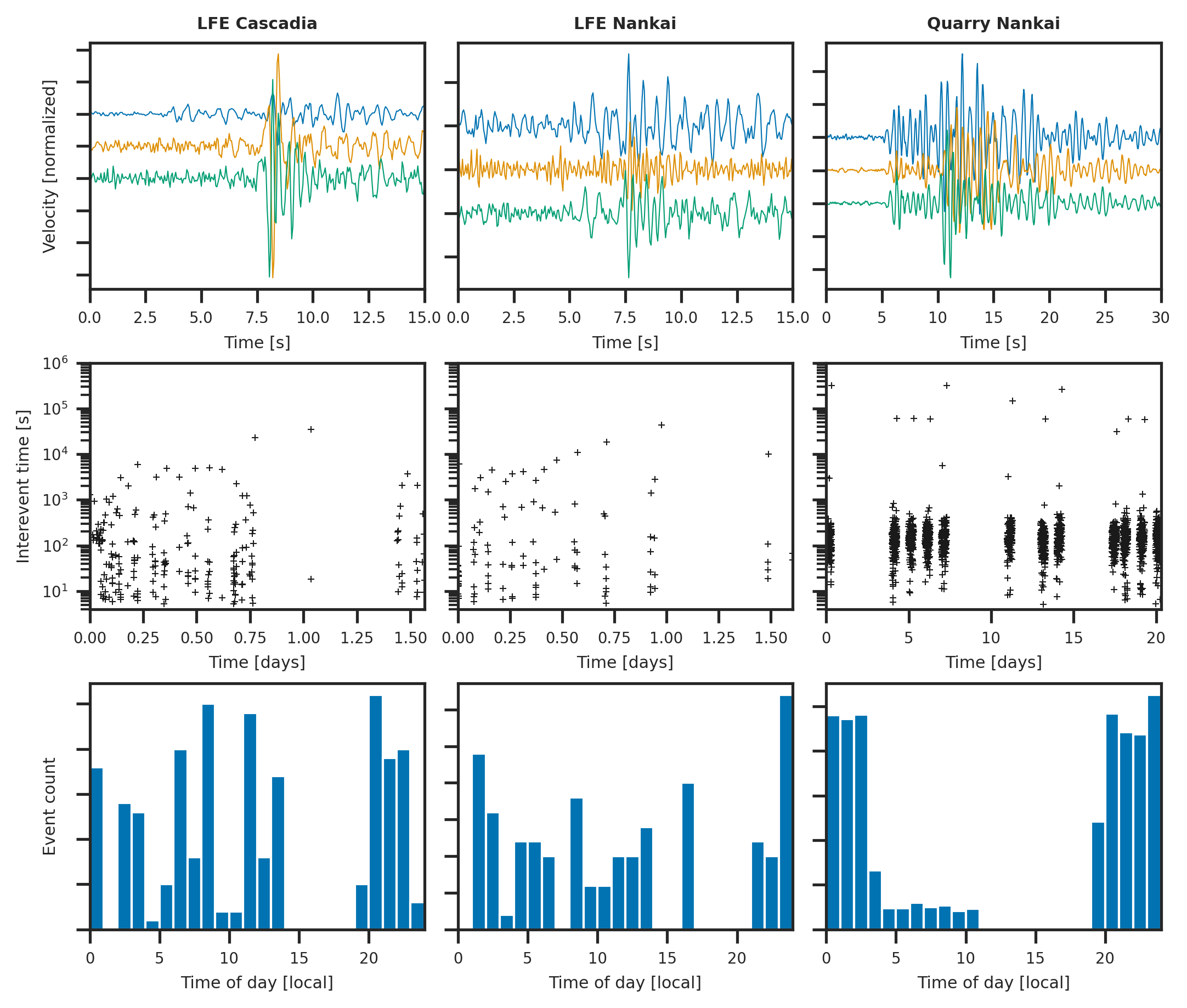}
\caption{Detections from the reference regions after template matching. The left two columns are LFE families from Cascadia and Nankai. The right column shows blasting in the Torigatayama limestone quarry. For a full description of the figure, see the caption of Figure~\ref{fig:lfe_examples}.}
\label{fig:lfe_examples_reference}
\end{figure*}

\begin{figure*}[ht!]
\centering
\includegraphics[width=\textwidth]{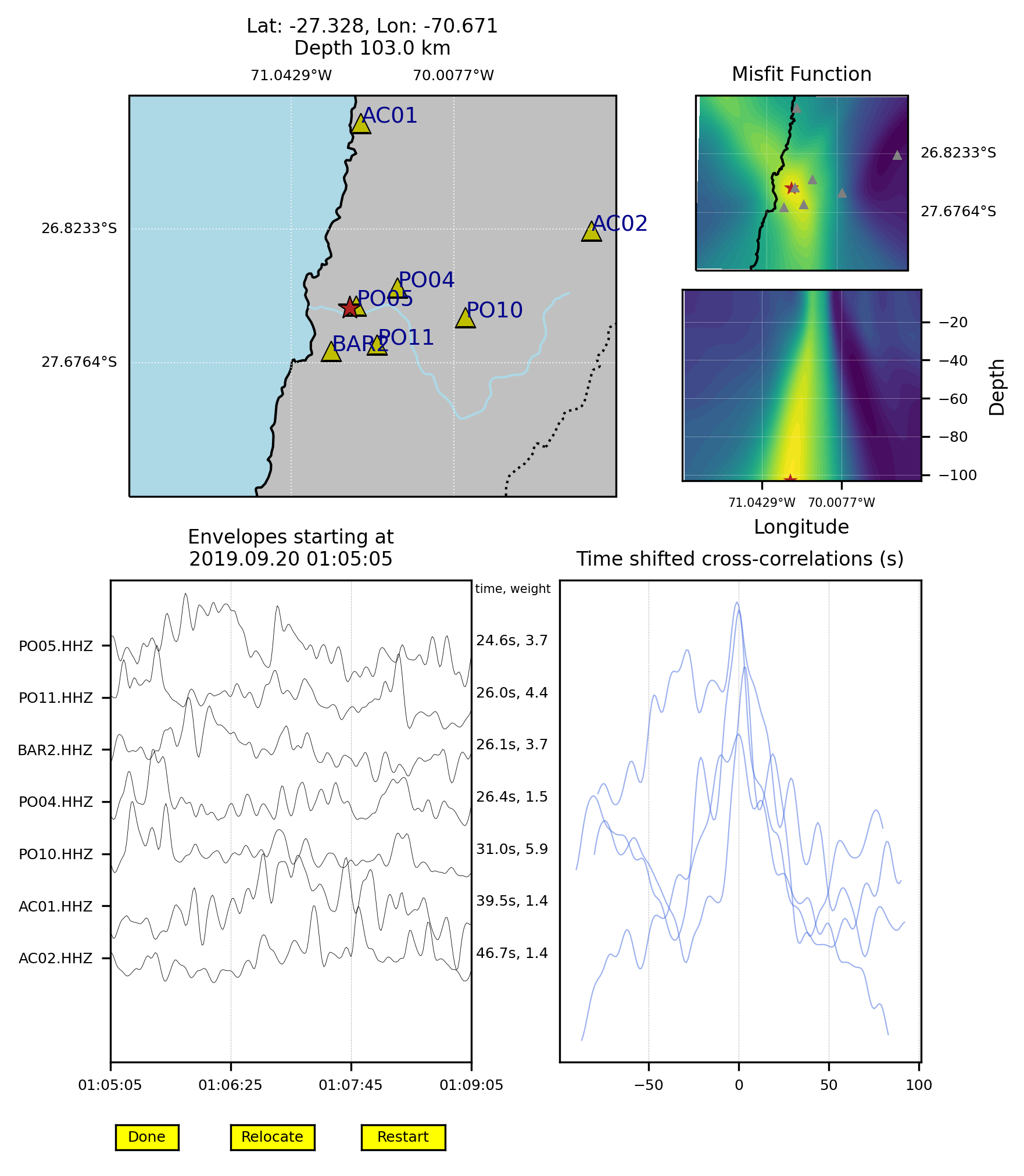}
\caption{Envelope correlation result for a detection from \cite{pasten-arayaAlongDipSegmentationSlip2022}. The detection occurs at the same time as listed in the original publication, but we reprocessed the data with an adjusted preprocessing and a 3D velocity model. The location suggests a deep or teleseismic source. The figure is based on the work of \citeA{wechAutomatedDetectionLocation2008}.}
\label{fig:pasten_enveloc1}
\end{figure*}

\begin{figure*}[ht!]
\centering
\includegraphics[width=\textwidth]{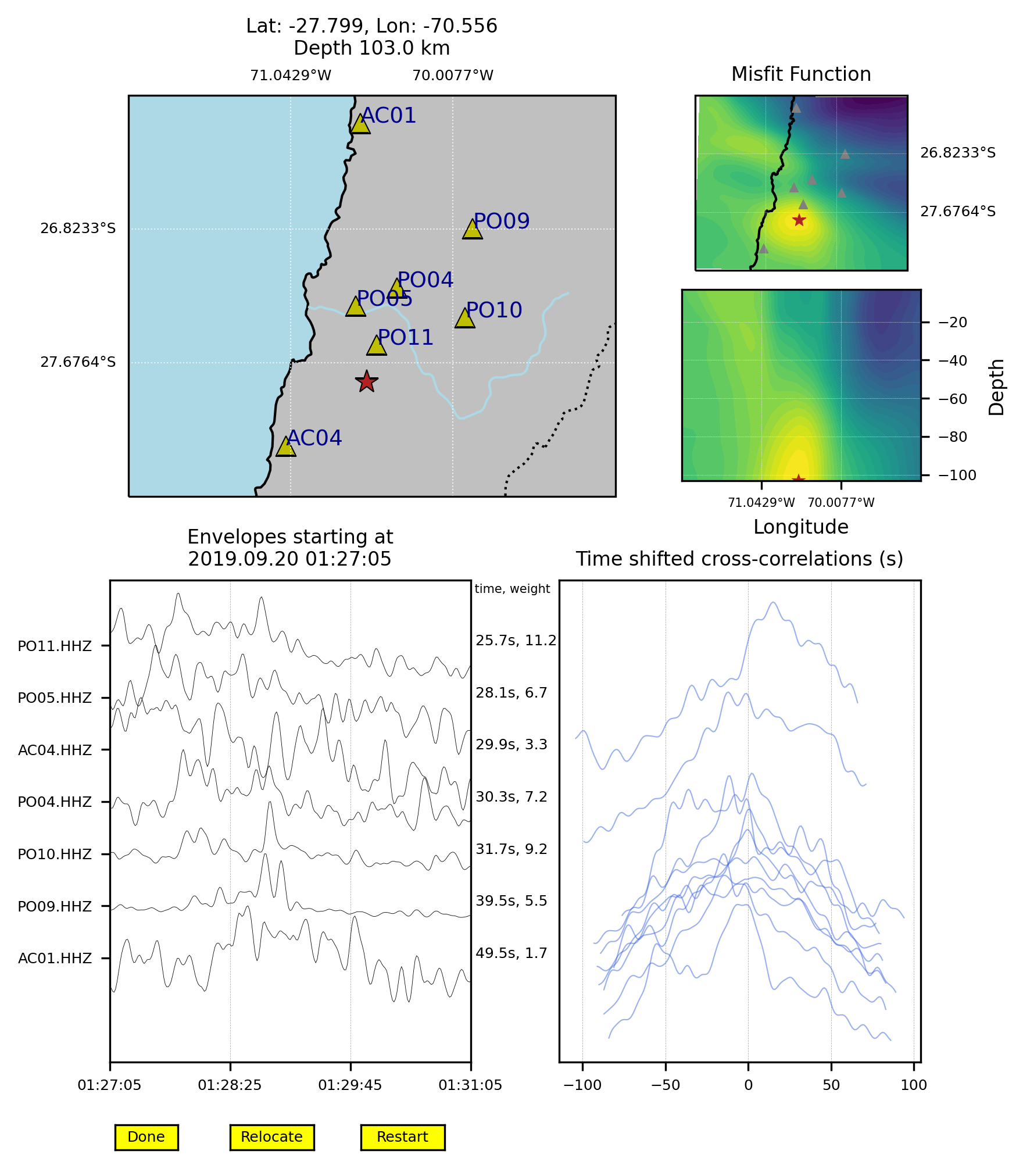}
\caption{Envelope correlation result for a detection from \cite{pasten-arayaAlongDipSegmentationSlip2022}. The detection occurs at the same time as listed in the original publication, but we reprocessed the data with an adjusted preprocessing and a 3D velocity model. The location suggests a deep or teleseismic source. The figure is based on the work of \citeA{wechAutomatedDetectionLocation2008}.}
\label{fig:pasten_enveloc2}
\end{figure*}

\begin{figure*}[ht!]
\centering
\includegraphics[width=\textwidth]{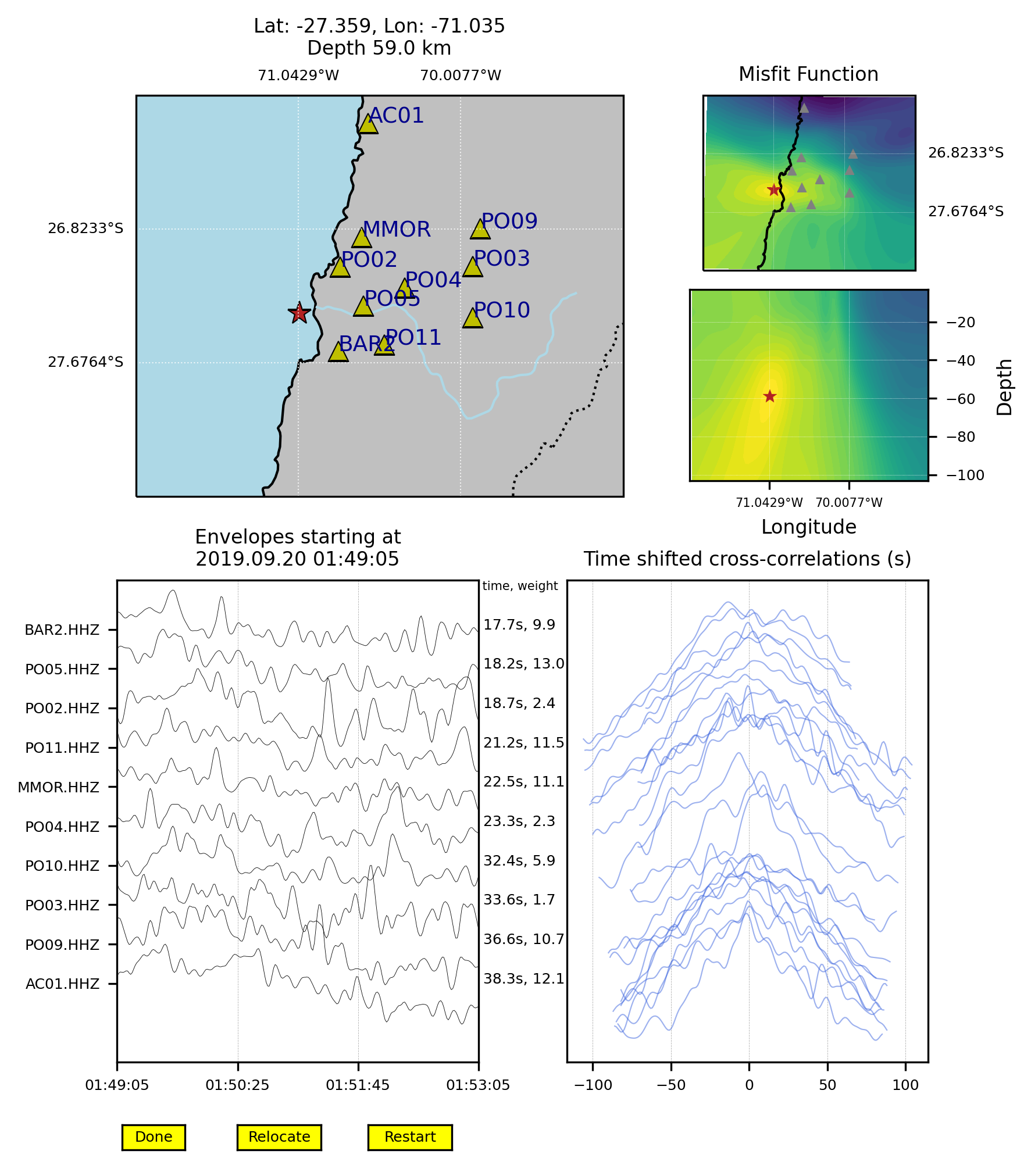}
\caption{Envelope correlation result for a detection from \cite{pasten-arayaAlongDipSegmentationSlip2022}. The detection occurs at the same time as listed in the original publication, but we reprocessed the data with an adjusted preprocessing and a 3D velocity model. The location suggests a deep or teleseismic source. The figure is based on the work of \citeA{wechAutomatedDetectionLocation2008}.}
\label{fig:pasten_enveloc3}
\end{figure*}

\begin{figure*}[ht!]
\centering
\includegraphics[width=\textwidth]{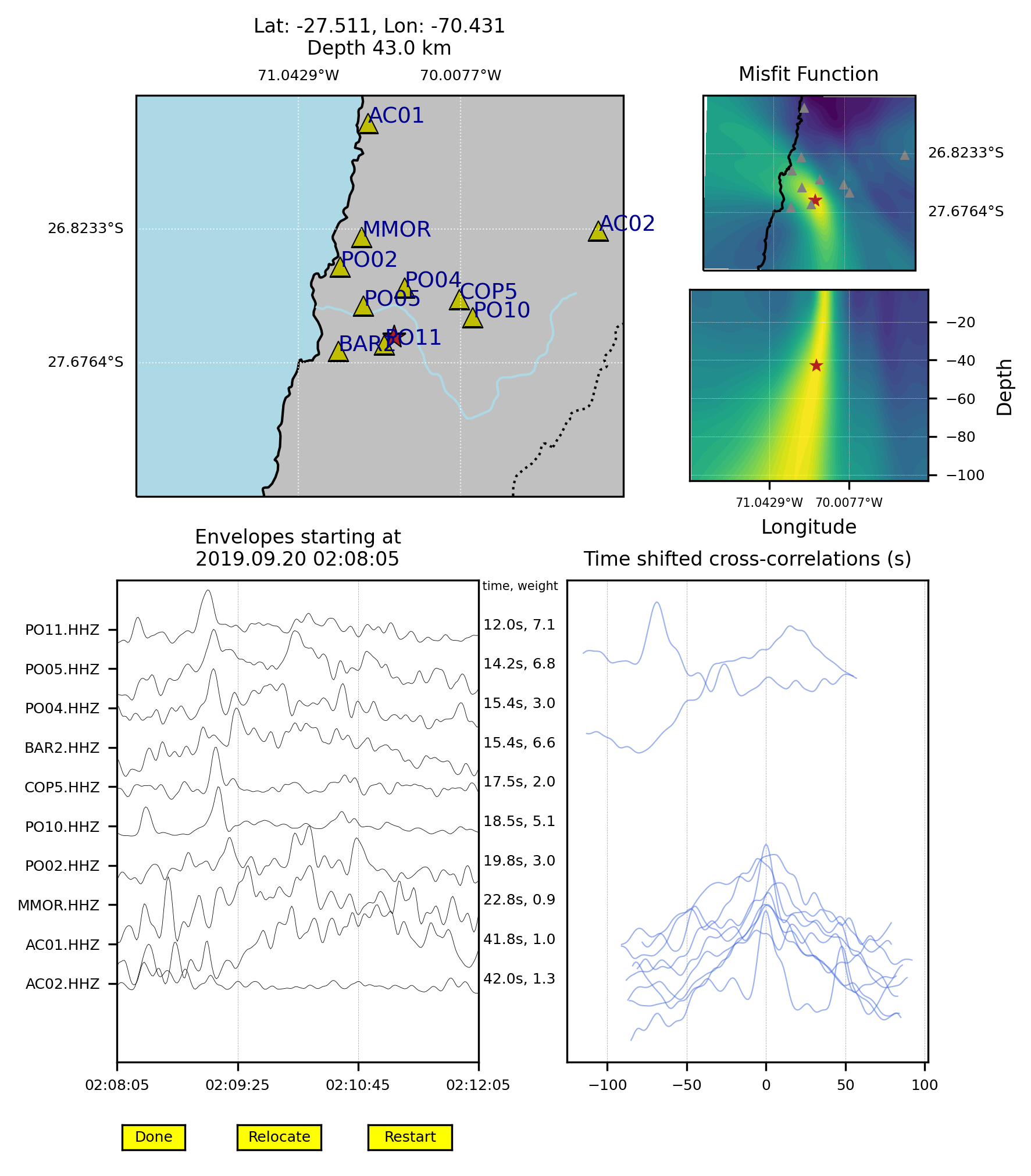}
\caption{Envelope correlation result for a detection from \cite{pasten-arayaAlongDipSegmentationSlip2022}. The detection occurs at the same time as listed in the original publication, but we reprocessed the data with an adjusted preprocessing and a 3D velocity model. The depth of this detection is purely constrained and is compatible with either local or teleseismic sources. The figure is based on the work of \citeA{wechAutomatedDetectionLocation2008}.}
\label{fig:pasten_enveloc4}
\end{figure*}

\begin{figure*}[ht!]
\centering
\includegraphics[width=\textwidth]{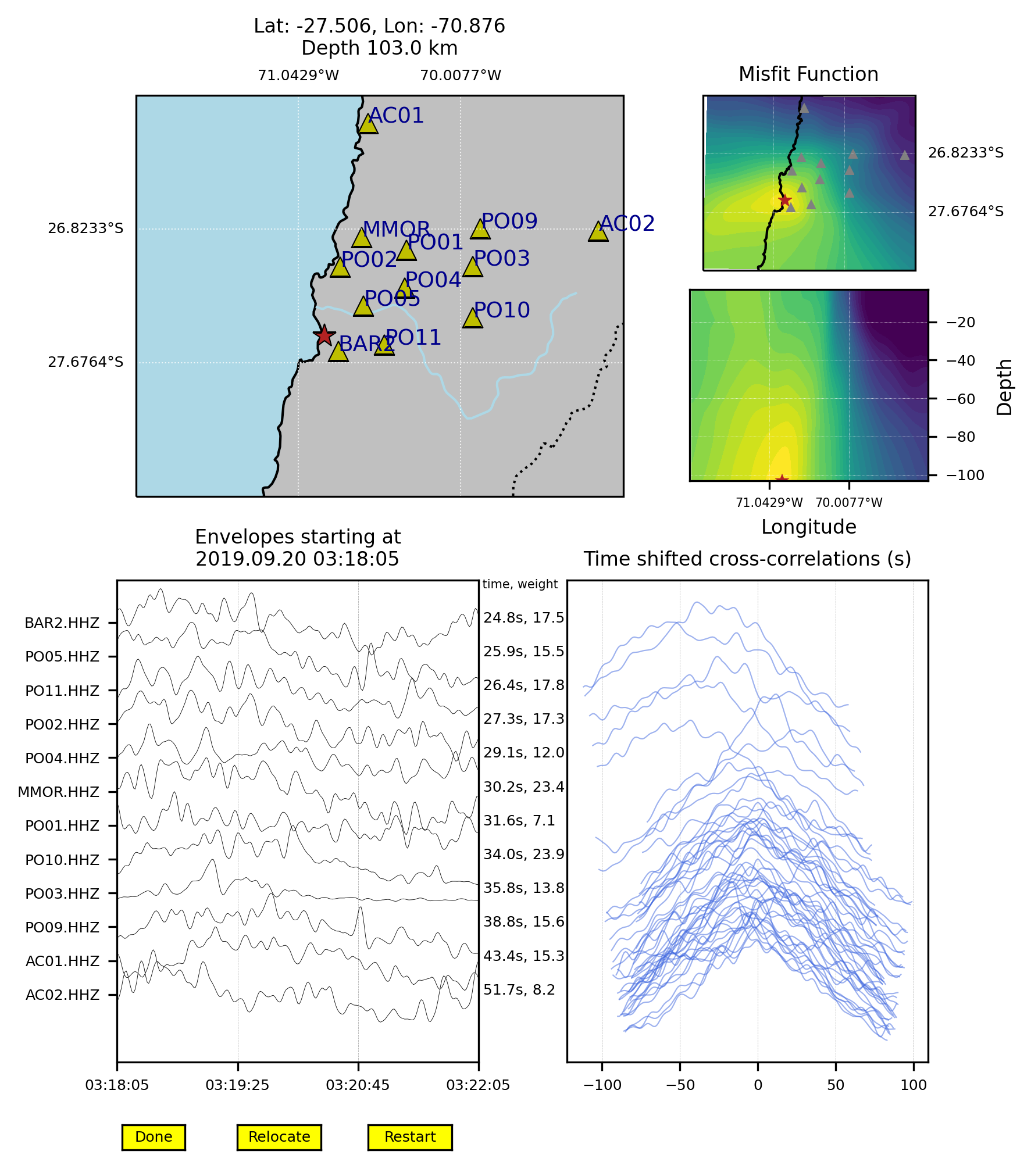}
\caption{Envelope correlation result for a detection from \cite{pasten-arayaAlongDipSegmentationSlip2022}. The detection occurs at the same time as listed in the original publication, but we reprocessed the data with an adjusted preprocessing and a 3D velocity model. The location suggests a deep or teleseismic source. The figure is based on the work of \citeA{wechAutomatedDetectionLocation2008}.}
\label{fig:pasten_enveloc5}
\end{figure*}

\begin{figure*}[ht!]
\centering
\includegraphics[width=\textwidth]{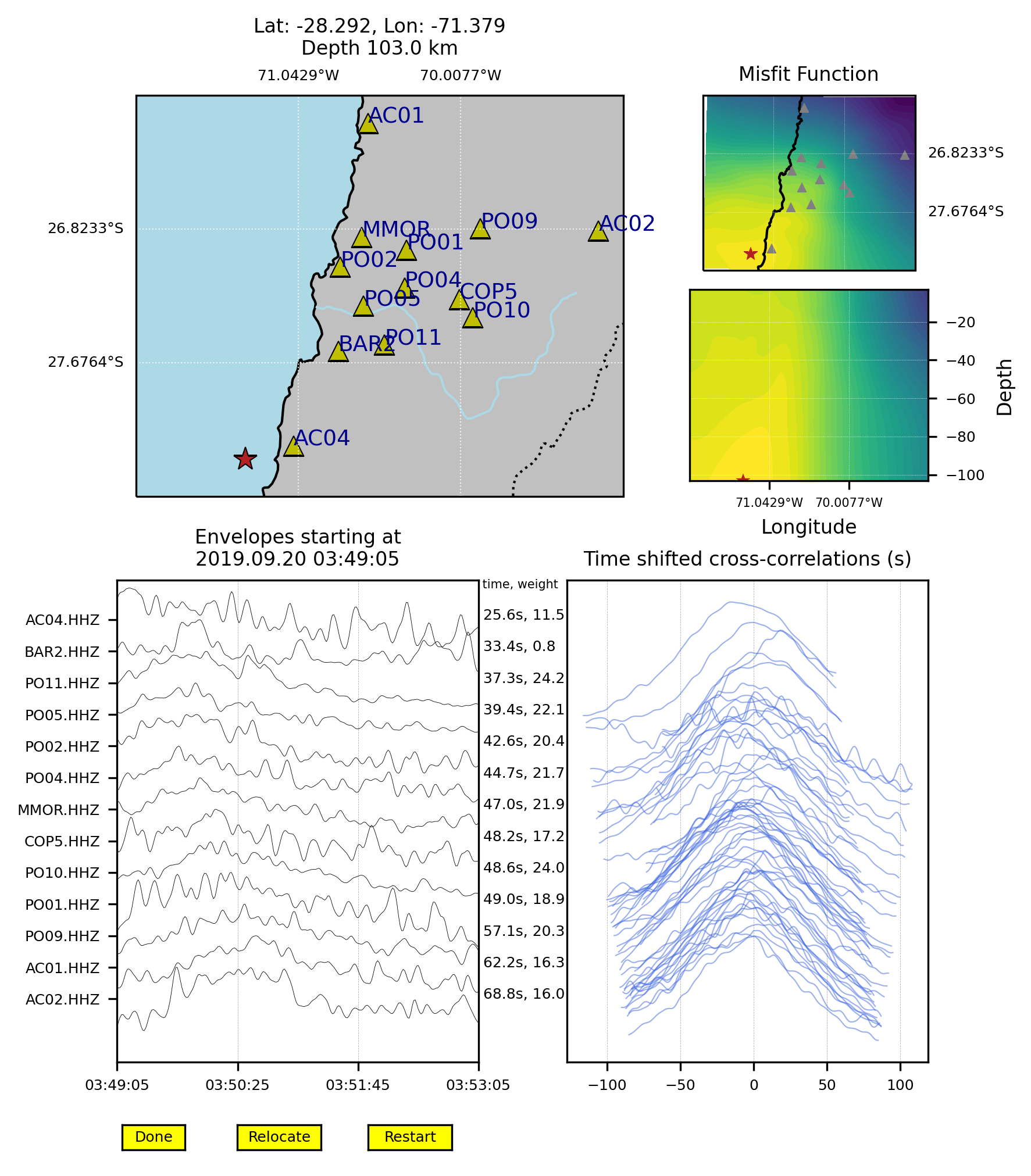}
\caption{Envelope correlation result for a detection from \cite{pasten-arayaAlongDipSegmentationSlip2022}. The detection occurs at the same time as listed in the original publication, but we reprocessed the data with an adjusted preprocessing and a 3D velocity model. The location suggests a deep or teleseismic source. The figure is based on the work of \citeA{wechAutomatedDetectionLocation2008}.}
\label{fig:pasten_enveloc6}
\end{figure*}

\begin{figure*}[ht!]
\centering
\includegraphics[width=0.8\textwidth]{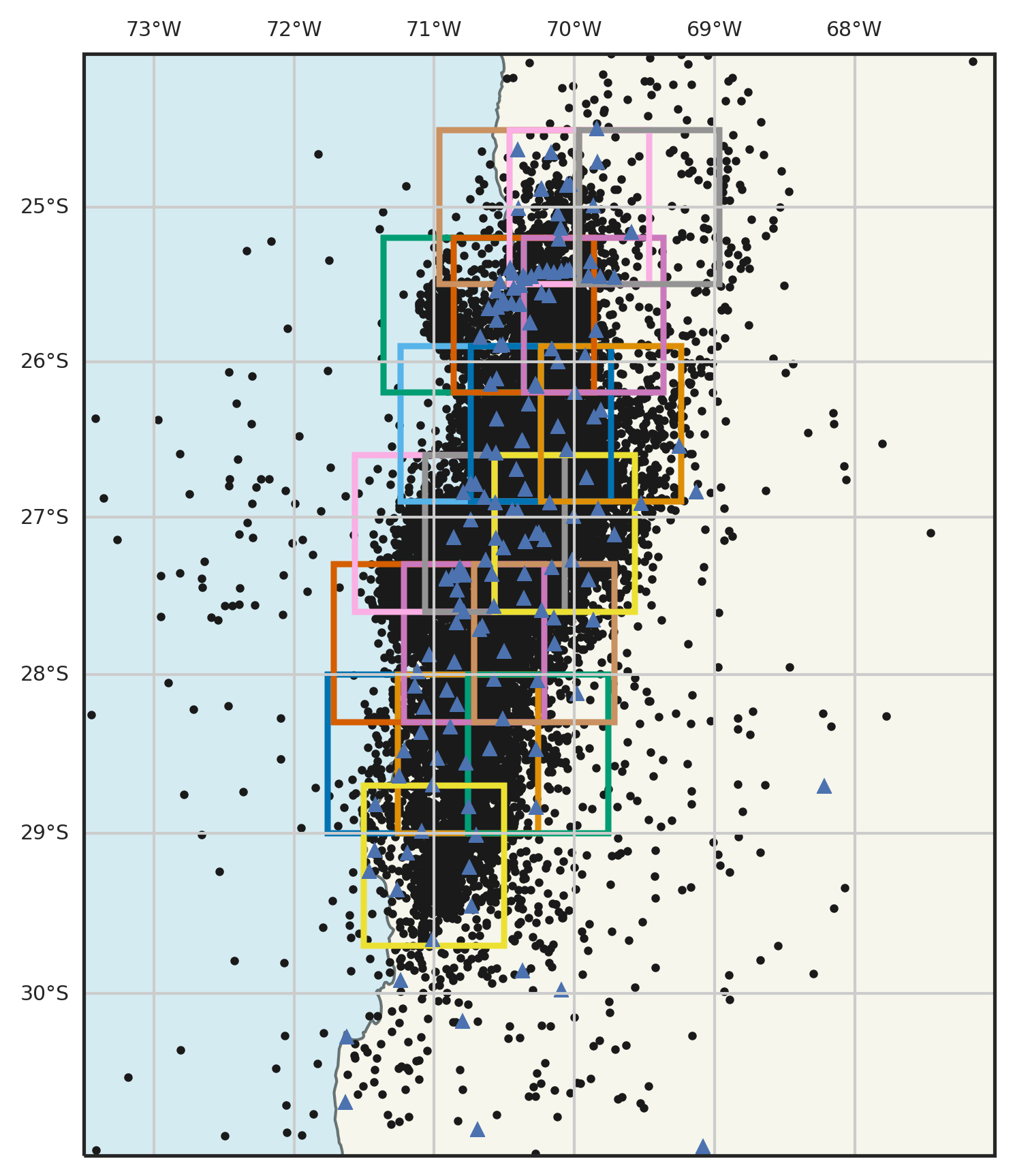}
\caption{Subdivision of the Chile study area into 19 overlapping cells of 1\degree by 1\degree for LFE template matching. Each cell is shown through its colored outline. Black dots denote initial LFE candidates. Blue triangle denote seismic stations.}
\label{fig:lfe_cells}
\end{figure*}

\end{document}